\documentclass[aps,preprint,prd,longbibliography,amsmath,superscriptaddress,amssymb,amsfonts]{revtex4-1}
\usepackage[utf8]{inputenc}
\usepackage[T1]{fontenc}
\usepackage[english]{babel}
\usepackage{csquotes}
\usepackage{mathtools}
\usepackage{graphicx}
\usepackage{subfigure}%
\usepackage{float}
\usepackage{booktabs}
\usepackage{mathrsfs}
\usepackage{tensor}
\usepackage{verbatim}
\usepackage{SIunits}
\usepackage{color}
\usepackage[colorlinks=true,
linkcolor=blue, 
anchorcolor=blue, 
citecolor=blue, 
urlcolor=blue
]{hyperref}

\date{\today}
\begin{document}
	\title{The Hall effects of vortex light in optical materials}
	\author{Wei-Si Qiu}
	\email{qiuws@mail2.sysu.edu.cn}
	\affiliation{School of Physics and Astronomy, Sun Yat-sen University, 519082 Zhuhai, China}
	\author{Li-Li Yang}
	\email{yanglli6@mail2.sysu.edu.cn}
	\affiliation{School of Physics and Astronomy, Sun Yat-sen University, 519082 Zhuhai, China}
	\author{Dan-Dan Lian}
	\email{liandd@mail.sysu.edu.cn}
	\affiliation{School of Physics and Astronomy, Sun Yat-sen University, 519082 Zhuhai, China}
	\author{Peng-Ming Zhang}
	\email{zhangpm5@mail.sysu.edu.cn}
	\affiliation{School of Physics and Astronomy, Sun Yat-sen University, 519082 Zhuhai, China}	
    \affiliation{Guangdong Provincial Key Laboratory
of Quantum Metrology and Sensing, Sun Yat-Sen University, Zhuhai 519082, China}
    \affiliation{Lanzhou Center for Theoretical Physics, Key Laboratory of Quantum Theory and Applications of MoE, and Key Laboratory of Theoretical Physics of Gansu Province, Lanzhou University, Lanzhou, Gansu 730000, China
}
   
\begin{abstract}

For light, its spin can be independent of the spatial distribution of its wave function, whereas its intrinsic orbital angular momentum does depend on this distribution. This difference suggests that the spin Hall effect might differ from the orbital Hall effect as light propagates through optical materials. In this paper, we model optical materials as curved space-time and investigate light propagation in two specific materials by solving the covariant Maxwell equations. We find that the trajectory of light with spin $\sigma$ and intrinsic orbital angular momentum $\ell$ deviates from that of light without angular momentum ($\sigma=0$ and $\ell=0$) by an angle $\theta_{\sigma,\ell} \propto 2\sigma+\ell$. In particular, the contribution of spin $\sigma$ to angle $\theta_{\sigma,\ell}$ is twice that of the intrinsic orbital angular momentum $\ell$, highlighting their differing effects on light propagation in optical materials. Furthermore, this angle $\theta_{\sigma,\ell}$ could potentially be observed experimentally, enhancing our understanding of the role of angular momentum in light propagation.

\end{abstract}	
	\maketitle
	\newpage
	\section{Introduction}\label{introduction}

The spin Hall effect (SHE) of light, arising from spin-orbit coupling in specific media or external fields, results in spin-dependent ray trajectories. This phenomenon is well documented in both theory and experiments \cite{bliokh2009spin,bliokh2009geometrodynamics,Bliokh_2006,Mieling_2023,fu2019spin,sheng2023photonic,ling2017recent,hosten2008observation,bliokh2008geometrodynamics,Bliokh_2015,PhysRevA.84.043806,PhysRevLett.102.123903,Duval:2018hzh,Duval:2012ye,Duval:2005ry,Duval:2005ky}. Similarly, a comparable effect occurs for light that contains intrinsic orbital angular momentum, known as the orbital Hall effect (OHE) \cite{beijersbergen1993astigmatic,bliokh2017theory}. It is generally believed that these two effects should be analogous in optical materials, due to similar Berry connection structures for spin and intrinsic orbital angular momentum \cite{bliokh2017theory,bliokh2009spin,PhysRevA.97.033843,PhysRevA.92.043805,Bliokh_2006}.

However, these two types of angular momentum exhibit significant distinctions. For light, its spin, characterized by a polarization vector \(\vec{\epsilon}\), can be independent of the spatial distribution of its wave function. The spin retains its physical meaning even when light is described as a point-like particle. In contrast, for vortex light, the intrinsic orbital angular momentum is determined by the phase factor \(e^{i\ell \phi}\), where \(\phi\) is the azimuthal angle. This phase factor depends on the spatial distribution. Consequently, vortex light cannot be treated as point-like particles; otherwise, the phase factor becomes undefined, and the intrinsic orbital angular momentum may lose its physical meaning.

Several studies have highlighted the distinctions between spin and intrinsic orbital angular momentum in specific contexts, leading to differences between the SHE and the OHE. For instance, these differences can manifest as a pure geometric shift in the centroid of light intensity (energy flux) \cite{wang2019anomalous}. In titanium (Ti), the electron orbital Hall conductivity is two orders of magnitude greater than the spin Hall conductivity, attributed to significant orbital texture and a larger orbital Berry curvature \cite{choi2023observation}. In curved spacetime, the relationships between intrinsic orbital angular momentum-dependent and spin-dependent separations and trajectories can differ markedly \cite{Qiu:2024rsr}.

The aforementioned differences between spin and intrinsic angular momentum suggest that the Hall effects associated with spin may differ from those induced by intrinsic orbital angular momentum. Therefore, this work focuses on investigating the spin and orbital Hall effects for light propagating within materials, aiming to identify the distinctions between spin and intrinsic orbital angular momentum. In this paper, the motion of light in a specific material can be effectively described as its motion in curved spacetime, characterized by a particular metric. This approach has been widely discussed in numerous studies \cite{BIALYNICKIBIRULA1996245,narimanov2009optical,genov2009mimicking,cheng2010omnidirectional,yang2012electromagnetic,sheng2013trapping,Duval:2005ky}. In this work, we use the energy-momentum tensor of light to describe its motion in materials. This method has been successfully applied to describe the motion of particles in the curved spacetime \cite{PhysRevD.105.104008,lian2024motion,Qiu:2024rsr}.

The structure of this paper is organized as follows: Section \ref{Dynamics of vortex wave packets in the equivalent optical material} introduces the dynamics of vortex light by numerically solving the covariant Maxwell equations within the effective metric for optical materials. In Section \ref{Physical system}, we construct a vortex Laguerre-Gaussian electromagnetic wave packet in flat spacetime and apply the center of its energy density to describe its motion. We successfully identify the distinctions between the SHE and the OHE. In Section \ref{The observation of the OHE in two realistic optical materials}, we select two realistic optical materials as examples and compute the transverse trajectory and deviation angle associated with the Hall effects. Section \ref{Discussion} presents a discussion of our findings.

In this work, we utilize the metric signature $(-,+,+,+)$ and adopt relativistic units where $c=\hbar=1$, with $c$ representing the speed of light in a vacuum. Greek indices span the four coordinates in a general coordinate system, while Latin indices $i, j, k, \ldots$ are confined to the three spatial coordinates. The Riemann curvature tensor is given by 
$R^\alpha_{\ \mu \nu \beta} = \partial_{\beta} \Gamma_{\mu \nu}^{\alpha} - \partial_{\nu} \Gamma_{\mu \beta}^{\alpha} + \Gamma_{\mu \nu}^{\gamma} \Gamma_{\beta \gamma}^{\alpha} - \Gamma_{\mu \beta}^{\gamma} \Gamma_{\nu \gamma}^{\alpha},$
where $\Gamma_{\mu \nu}^{\alpha}$ denotes the Christoffel symbols.

\section{Dynamics of vortex light in optical materials}\label{Dynamics of vortex wave packets in the equivalent optical material}

As introduced in Section \ref{introduction}, we can effectively model the behavior of vortex light in inhomogeneous media by treating these media as curved spacetimes. The dynamics of vortex light are governed by the source-free Maxwell equations. Within a gravitational field, these equations are expressed as:
\begin{equation}\label{maxwell}
	\nabla_\nu F^{\mu\nu} = 0,
\end{equation}
where \(F^{\mu \nu} = \nabla^\mu A^\nu - \nabla^\nu A^\mu\) is the covariant electromagnetic field tensor, and \(\nabla_\nu\) denotes the covariant derivative. When considering the motion of light in a gravitational field and employing the Lorenz condition \(\nabla_\mu A^\mu = 0\), Eq. \eqref{maxwell} can be simplified to:
\begin{equation}\label{eom}
	\nabla^\mu \nabla_\mu A^\nu = -g^{\rho \nu} R_{\mu \rho} A^{\mu},
\end{equation}
where \(R_{\mu \rho}\) represents the Ricci tensor.

A perfectly impedance-matched optical medium can be effectively modeled by a curved spacetime, where the metric \(g_{\mu\nu}\) is described as follows \cite{PhysRev.118.1396,carini1992phase,wu2022testing,chanda2019jacobi,gibbons2019gravitational,fernandez2016anisotropic}:
\begin{align}\label{tensorial permittivity,tensorial permeability}
	\sqrt{-g} \frac{g^{ij}}{g_{00}} = -\epsilon^{ij} = -\mu^{ij}, \quad \frac{g_{0i}}{g_{00}} = -\alpha_{i},
\end{align}
where \(\epsilon_{ij}\) and \(\mu_{ij}\) represent the tensorial permittivity and permeability, respectively, and \(\alpha_i\) denotes the magnetoelectric coupling vector. Specifically, we define the medium's refractive index as \(n = n(t,x,y,z)\) and its four-velocity as \(u_\alpha\). In this context, the metric for the effective curved spacetime takes the form:
\begin{equation}
	g_{\alpha \beta} = \eta_{\alpha \beta} + (1 - n^{-2}) u_{\alpha} u_{\beta}
\end{equation} 
\cite{PhysRevD.97.065001,gordon1923lichtfortpflanzung,PhysRevD.105.104061}, where \(\eta_{\alpha\beta} = \text{diag}\{-1,1,1,1\}\) is the Minkowski metric. Assuming a stationary medium with \(\partial_t n = 0\) and setting \(u^\alpha = (1,0,0,0)\), the above metric simplifies to:
\begin{equation}\label{metric1}
	g_{\alpha \beta} = \eta_{\alpha \beta} + (1 - n^{-2}) \delta_{0\alpha} \delta_{0\beta}.
\end{equation}
If we expand the refractive index to first-order approximation: \(n \simeq C_0 + \vec{C_1}(\vec{x}) \cdot \vec{x}\), where \(C_0\) is a constant, we can express the metric (\ref{metric1}) as:
\begin{equation}\label{metric2}
	g_{\alpha \beta} = \eta'_{\alpha \beta} + h_{\alpha \beta} = \eta'_{\alpha \beta} + h_{00} \delta_{0\alpha} \delta_{0\beta}.
\end{equation}
In our analysis, to simplify the calculations of the Maxwell equations (\ref{eom}), we utilize weak field approximations. Thus, \(|h_{00}| \ll 1\) is always satisfied.

Accordingly, it is practical to consider the gravitational impact on the electromagnetic field \(A^\nu\) in Eqs.~(\ref{eom}) as a perturbative effect:
\begin{equation}
	A^\nu = \bar{A}^\nu + \widetilde{A}^\nu,
\end{equation}
with \(\bar{A}^\nu\) representing the zeroth-order term of the gravitational potential, satisfying the equation \(\eta'^{\rho\mu} \partial_\rho \partial_\mu \bar{A}^\nu = 0\). According to Eq. \eqref{eom}, the perturbation \(\widetilde{A}^\nu\), being first order, can be approximately described by:
\begin{equation}\label{eom-first}
	\eta'^{\mu \rho} \partial_\mu \partial_\rho \widetilde{A}^\nu \simeq h^{\mu \rho} \partial_\mu \partial_\rho \bar{A}^\nu + \eta'^{\mu\rho} \widetilde{\Gamma}^\sigma_{\mu\rho} \partial_\sigma \bar{A}^\nu - \eta'^{\mu \rho} \partial_\rho \widetilde{\Gamma}^\nu_{\mu\sigma} \bar{A}^\sigma - 2 \eta'^{\mu \rho} \widetilde{\Gamma}^\nu_{\rho\sigma} \partial_\mu \bar{A}^\sigma - \eta'^{\rho \nu} \tilde{R}_{\mu \rho} \bar{A}^\mu,
\end{equation}
where \(\widetilde{\Gamma}_{\lambda\alpha}^{\beta} = \frac{1}{2} \eta'^{\sigma \gamma} (\partial_{\alpha} h_{\lambda \gamma} + \partial_{\lambda} h_{\alpha \gamma} - \partial_{\gamma} h_{\lambda \alpha})\) denotes the first-order terms of the affine connection, and \(\tilde{R}_{\mu \rho}\) is the first order of \(R_{\mu \rho}\). By expanding the Christoffel symbols and the Riemann tensor, we can simplify the equations to:
\begin{equation}\label{final equations}
	\eta'^{\mu \rho} \partial_\mu \partial_\rho \widetilde{A}^\nu \simeq h^{00} \partial_{t}^2 \bar{A}^\nu - \frac{1}{2} \eta'^{00} \eta'^{\sigma \xi} \partial_{\xi} h_{00} \partial_{\sigma} \bar{A}^{\nu} - \eta'^{00} \eta'^{\nu 0} \partial_{0} \bar{A}^{i} \partial_{i} h_{00} + \frac{1}{2} (\partial_{i} \partial_{j} h_{00}) \eta'^{00} \eta'^{i \nu} \bar{A}^{j}.
\end{equation}

\section{Physical system and the initial state}\label{Physical system}

In the absence of gravity within flat spacetime, vortex light can be characterized by a vortex Laguerre-Gaussian electromagnetic wave packet \cite{liu2023threshold,Qiu:2024rsr}. The expression for this wave packet in momentum space is given by:
\begin{equation}\label{LG-wave}
	\bar{A}_{f}^i(\vec{k},t) = \frac{\sqrt{2} N (\sigma_{\perp} k_{\perp})^{\ell}}{\omega^{3/2}} \epsilon^i \exp \left(-\frac{k_{\perp}^{2} \sigma_{\perp}^{2}}{2} - \frac{(k_{z} - p_{z})^{2} \sigma_{z}^{2}}{2} + i \ell \phi_{k} - i\omega t\right),
\end{equation}
where \(\vec{\epsilon} = \{\epsilon^i\} = (k_{z}^{2} + k_{y}^{2} - i \sigma k_{x} k_{y}, i \sigma k_{z}^{2} + i \sigma k_{x}^{2} - k_{x} k_{y}, -i \sigma k_{y} k_{z} - k_{x} k_{z})^\text{T}\), with \(\sigma = 0, \pm 1\), represents the polarization vector. The wave packet adheres to the Coulomb Gauge, \(\partial_i \bar{A}^i = 0\), with \(\bar{A}^0 = 0\). This condition ensures that our wave function satisfies Maxwell's equations: \(\eta'^{\mu \rho} \partial_\mu \partial_\rho \bar{A}^v = 0\), which is derived from Eq.~\eqref{eom} in flat spacetime. The transverse component of the wave vector is denoted as \(k_{\perp} = \sqrt{k_x^2 + k_y^2}\). The angular frequency, \(\omega\), is defined as \(\omega = \frac{1}{n}\sqrt{k_x^2 + k_y^2 + k_z^2}\), and the azimuthal angle, \(\phi_k\), is given by \(\phi_k = \arctan\left(\frac{k_y}{k_x}\right)\). The parameter \(p_z\) refers to the average initial momentum of the wave packet along the \(z\)-axis, and \(N\) is the normalization factor. Moreover, it can be demonstrated that the average spin, intrinsic orbital angular momentum, and momentum of this vortex wave packet are equal to \(\sigma\), \(\ell\), and \(p_z\), respectively \cite{Qiu:2024rsr}.

The wave packet is expressed in momentum space, necessitating the transformation of Eq. \eqref{eom-first} into momentum space through a Fourier Transform across three-dimensional space. Given the wave packet \(\bar{A}_f^i\) as depicted in Eq. \eqref{LG-wave}, the momentum space representation of Eq. \eqref{eom-first} can be expressed as:
\begin{equation}\label{eom-k}
	\left(\partial_{t}^{2} + \omega^{2}\right) \tilde{A}_{f}^{\rho} = a_1^{\rho}(\vec{k}) e^{-i \omega t} + a_2^{\rho}(\vec{k}) e^{-i \omega t} t,
\end{equation}
where \(\widetilde{A}^\rho_f = \frac{1}{2\pi^3} \int \widetilde{A}^\rho \exp(-i\vec{k} \cdot \vec{x}) \text{d}^3x\) represents the perturbation term in momentum space. The coefficient \(a_1^{\rho}(\vec{k})\) encompasses all factors associated with terms that include the time-dependent factor \(\exp(-i\omega t)\). Similarly, the coefficient \(a_2^{\rho}(\vec{k})\) encapsulates all factors related to terms that incorporate the time-dependent expression \(t \exp(-i\omega t)\). Given the initial conditions:
\begin{equation}\label{initial conditions}
	\left.\tilde{A}_{f}^{\rho}\right|_{t=0} = 0 \quad \text{and} \quad \left.\partial_{t} \tilde{A}_{f}^{\rho}\right|_{t=0} = 0.
\end{equation}
Eq. \eqref{eom-k} can be analytically solved. The perturbation term \(\widetilde{A}^\rho_f\) is expressed as:
\begin{equation}\label{general solution}
	\tilde{A}_{f}^{\rho} = \frac{a_1^{\rho}(\vec{k})(1 - e^{2 i {\omega} t} + 2 i {\omega} t) e^{-i {\omega} t}}{4 {\omega}^{2}} + \frac{a_2^{\rho}(\vec{k})(2 {\omega} t - i + 2 i {\omega}^{2} t^{2} + i e^{2 i {\omega} t}) e^{-i {\omega} t}}{8 {\omega}^{3}}.
\end{equation}
Accordingly, the wave function of the free-falling vortex wave packet in coordinate space is determined through the inverse Fourier Transform:
\begin{equation}
	A^{\rho} = \int A^{\rho}_f \exp(i \vec{k} \cdot \vec{x}) \, \text{d}^{3}k, \quad A^{\rho}_f = \bar{A}^{\rho}_f + \widetilde{A}^{\rho}_f.
\end{equation}

Unlike the point-like particle, the wave packet is not local and we can use the center of the symmetric energy-momentum tensor: $T_{s}^{\mu \nu}=F^{\mu \rho} F_{\rho}^{\nu}-\frac{1}{4} g^{\mu \nu} F^{\lambda \rho} F_{\lambda \rho}$ to describe the evolution of the wave packet \cite{PhysRevD.105.104008,Qiu:2024rsr}:
\begin{equation}\label{center}
	\left\langle x^{i}\right\rangle^{\mu \nu}=\frac{\int \sqrt{-g} x^{i} T_s^{\mu \nu} \mathrm{d}^{3} x}{\int \sqrt{-g} T_s^{\mu \nu} \mathrm{d}^{3} x},
\end{equation}
Under the weak field approximation, they can always be expressed as terms of the zeroth and the first order of $h_{\mu \nu}$: $T_s^{\mu \nu}\simeq \bar{T}_s^{\mu \nu}+\widetilde{T}_s^{\mu \nu}$. Utilize the energy density component:  $T^{00}_s$, and (\ref{center}) becomes:
\begin{equation}\label{ed-center}
	\langle x^i\rangle _{se}\simeq \frac{\int x^i(\widetilde{T}^{00}_s+\bar{T}_s^{00}) \text{d}^3x}{\int \bar{T}^{00}_s \text{d}^3x}.
\end{equation} 
The evolutions along $x$ and $z$ directions are mainly governed by the geodesic line, and therefore, we will mostly consider the transverse trajectory:$\langle y\rangle _{se}$.

\subsection{The angular-momentum dependent transverse deviations}

\begin{figure}
	\centering
	\includegraphics[width=0.7\linewidth]{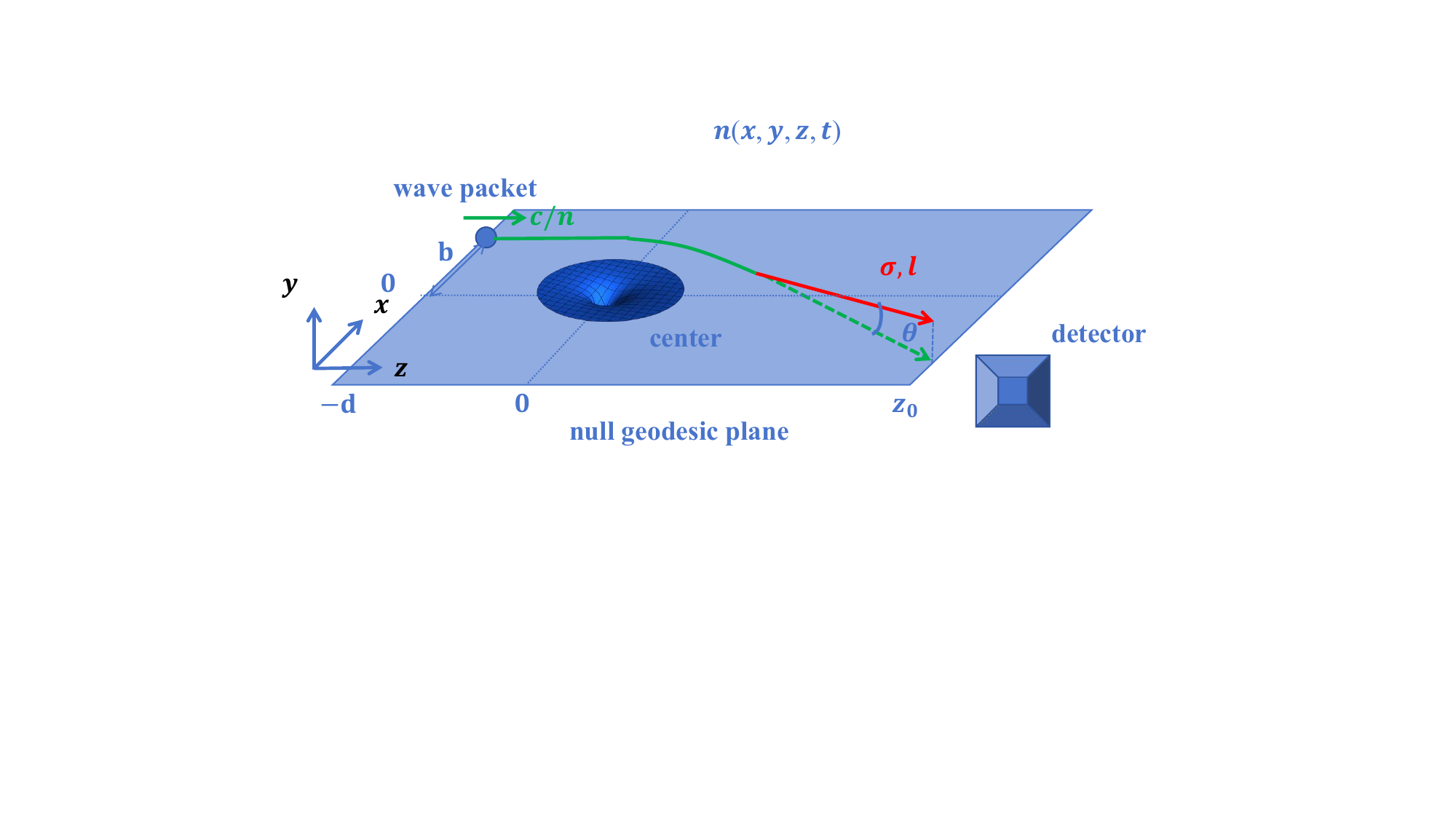}
	\caption{The motion of the wave packet within a material with finite boundaries and the subsequent detection of the deviation angle are considered. In the figure, both \(\sigma\) and \(\ell\) are assumed to be positive.}
	\label{fig:picture}
\end{figure}

As shown in Fig. \ref{fig:picture}, light possessing intrinsic angular momentum can exhibit a transverse deviation from the primary propagation plane that depends on the angular momentum. In the given geometric configuration, this transverse deviation occurs along the $y$-axis. In this work, the center of the wave packet is defined using energy density. The transverse velocity along the $y$-axis is given by:
\begin{equation}\label{analysis1}
	\frac{d\langle y \rangle_{\sigma,\ell}}{dt} = \frac{d}{dt}(\frac{\int y (\tilde{T}_s^{00}+\bar{T}_s^{00}) \mathrm{d}^{3} x} {\int \bar{T}_s^{00}\mathrm{d}^{3} x}).
\end{equation} 
Here, the expression $\int y \bar{T}^{00}_s \text{d}^3 x/\int \bar{T}^{00}_s\text{d}^3 x$ represents the center of the zeroth-order component of the wave packet, described by $\bar{A}^\mu$. According to Eq. \eqref{LG-wave}, this expression should vanish as the wave packet propagates within the two materials. Additionally, the definition $E\equiv \int \bar{T}^{00}_s\text{d}^3 x$ denotes the total energy of this zeroth-order component $\bar{A}^\mu$, which remains constant over time. Therefore, Eq.~\eqref{analysis1} simplifies to:
\begin{equation}\label{analysis2}
	\frac{d\langle y \rangle_{\sigma,\ell}}{dt} =\frac{1}{E}\frac{d}{dt}\int y \tilde{T}_s^{00} \mathrm{d}^{3}x.
\end{equation} 

In curved spacetime, we have the following equations:
\begin{equation}\label{conservation of flux}
	\nabla_\mu T_s^{\mu \nu} = 0,\quad \partial_\mu (x^\lambda T^{\mu\nu}_s)= T^{\lambda\nu}_s - x^\lambda\Gamma^{\mu}_{\mu\rho} T^{\rho\nu}_s - x^\lambda\Gamma^{\nu}_{\mu\rho}T^{\mu\rho}_s.
\end{equation}
Considering only the first-order terms of $h_{\mu \nu}$, we obtain:
\begin{align}
	&\partial_\mu (\tilde{T}_s^{\mu \nu}) = - \tilde{\Gamma}^\mu_{\mu \rho}  \bar{T}_s^{\rho \nu} - \tilde{\Gamma}^\nu_{\mu \rho} \bar{T}_s^{\mu \rho}.\\
	&\partial_\mu (x^\lambda\tilde{T}_s^{\mu \nu}) = \tilde{T}^{\lambda\nu}_s-x^\lambda \tilde{\Gamma}^\mu_{\mu \rho}  \bar{T}_s^{\rho \nu} - x^\lambda\tilde{\Gamma}^\nu_{\mu \rho} \bar{T}_s^{\mu \rho}.
\end{align}
After integrating over three-dimensional space, this equation becomes:
\begin{align}
	&\frac{d}{dt}\int \tilde{T}_s^{0\nu} \text{d}^3 x= - \int \tilde{\Gamma}^\mu_{\mu \rho}  \bar{T}_s^{\rho \nu} \text{d}^3 x- \int \tilde{\Gamma}^\nu_{\mu \rho} \bar{T}_s^{\mu \rho}\text{d}^3 x,\\
	&\frac{d}{dt}\int x^\lambda \tilde{T}_s^{0\nu} \text{d}^3 x=\int \tilde{T}^{\lambda\nu}_s \text{d}^3 x - \int x^\lambda \tilde{\Gamma}^\mu_{\mu \rho}  \bar{T}_s^{\rho \nu} \text{d}^3 x- \int x^\lambda\tilde{\Gamma}^\nu_{\mu \rho} \bar{T}_s^{\mu \rho}\text{d}^3 x,
\end{align}
where the surface integrals have been neglected. Specifically, we have the following relation:
\begin{align}
	&\frac{d}{dt}\int \tilde{T}_s^{02} \text{d}^3 x= - \int \tilde{\Gamma}^\mu_{\mu \rho}  \bar{T}_s^{\rho 2} \text{d}^3 x- \int \tilde{\Gamma}^2_{\mu \rho} \bar{T}_s^{\mu \rho}\text{d}^3 x,\\
	&\frac{d}{dt}\int y \tilde{T}_s^{00} \text{d}^3 x=\int \tilde{T}^{02}_s \text{d}^3 x - \int y \tilde{\Gamma}^\mu_{\mu \rho}  \bar{T}_s^{\rho 0} \text{d}^3 x- \int y\tilde{\Gamma}^0_{\mu \rho} \bar{T}_s^{\mu \rho}\text{d}^3 x.\label{eq-dydt}
\end{align}

Under the initial conditions specified by Eq.~\eqref{initial conditions}, the first term in Eq.~\eqref{eq-dydt} vanishes at \( t = 0 \). For \( t/b \ll 1 \), this term is expected to be proportional to time $t$, as \( \int \tilde{T}^{02}_s\text{d}^3 x \propto t \), indicating that it is of first order in \( O(t/b) \). In contrast, the remaining terms are of zeroth order in \( O(t/b) \). When retaining only the leading-order contributions of \( O(t/b) \) in Eq.~\eqref{eq-dydt}, the term \( \int \tilde{T}^{02}_s\text{d}^3 x \) can be neglected. By incorporating the metric defined in Eq.~\eqref{metric1}, the transverse velocity $d\langle y \rangle_{\sigma,\ell}/dt$ simplifies to:
\begin{equation}\label{analysis3}
	\frac{d\langle y \rangle_{\sigma,\ell}}{dt}\simeq-\frac{2}{E} \int y \tilde{\Gamma}^0_{0 \rho}  \bar{T}^{0\rho}_s \, \mathrm{d}^3 x.
\end{equation}

In this study, the radius of the wave packet is set to be much smaller than its distance from the center of the metric. The initial position of the wave packet is \(\vec{r}_0 = (b,0,0)\), while the center of the metric is located at \(\vec{r}_0 = (0, 0, 0)\). Consequently, we can expand the connection \(\Gamma^{\mu}_{\rho\lambda}\) within the wave packet as a series in \(\delta x^i/b\), where $\delta x^i=x^i-r_0^i$. By retaining only the first-order terms in \(\delta x^i/b\), we obtain:
\begin{equation}\label{analysis4}
	\frac{d\langle y \rangle_{\sigma,\ell}}{dt}\simeq-\frac{2}{E} \tilde{\Gamma}^0_{0 \rho} \Big|_{\vec{r}_0} \int \delta y \, \bar{T}^{0\rho}_s \, \mathrm{d}^3 x.
\end{equation}

The symmetric energy-momentum tensor $T_s^{\mu \nu}$ can be decomposed into two parts: the canonical energy-momentum tensor $T_c^{\mu \nu}$ and the Belinfante-Rosenfeld correction term \cite{peskin2018introduction}:
\begin{equation}
	T_s^{\mu \nu} = T_c^{\mu \nu} + \partial_\lambda B^{\lambda \mu \nu},
\end{equation}
where $B^{\lambda \mu \nu} \equiv \frac{1}{2}(S^{\mu \nu \lambda} + S^{\nu \mu \lambda} - S^{\lambda \nu \mu}) = -B^{\mu \lambda \nu}$, and $S^{\lambda \nu \mu}$ is the spin angular momentum tensor satisfying the antisymmetric relation $S^{\lambda \nu \mu} + S^{\lambda \mu \nu} = 0$. Therefore, Eq.~(\ref{analysis4}) can be rewritten as:
\begin{equation}\label{analysis5}
	\begin{split}
		\frac{d\langle y \rangle_{\sigma,\ell}}{dt}=&-\frac{2}{E} \tilde{\Gamma}^0_{0 \rho} \Big|_{\vec{r}_0} \int \delta y ( \bar{T}^{0\rho}_c + \partial_\lambda B^{\lambda 0 \rho} ) \mathrm{d}^3 x \\=& -\frac{2}{E} \tilde{\Gamma}^0_{0 \rho} \Big|_{\vec{r}_0} \int ( \delta y \, \bar{T}^{0\rho}_c + \partial_\lambda (\delta y \, B^{\lambda 0 \rho}) - B^{\lambda 0 \rho} \partial_\lambda \delta y ) \mathrm{d}^3 x.
	\end{split}
\end{equation}

The first part on the right-hand side of Eq.~(\ref{analysis5}) can be expanded as follows:
\begin{align}\label{expanded_first_term}
		-\frac{2}{E} \tilde{\Gamma}^0_{0 \rho} \Big|_{\vec{r}_0} \int \delta y \, \bar{T}^{0\rho}_c \mathrm{d}^3 x &=-\frac{2}{E} \left( \tilde{\Gamma}^0_{0 0} \Big|_{\vec{r}_0} \int \delta y \, \bar{T}^{00}_c \, \mathrm{d}^3 x + \tilde{\Gamma}^0_{0 1} \Big|_{\vec{r}_0} \int \delta y \, \bar{T}^{01}_c \, \mathrm{d}^3 x\right)\nonumber\\
		&-\frac{2}{E}\left( \tilde{\Gamma}^0_{0 2} \Big|_{\vec{r}_0} \int \delta y \, \bar{T}^{02}_c \,\mathrm{d}^3 x + \tilde{\Gamma}^0_{0 3} \Big|_{\vec{r}_0} \int \delta y \, \bar{T}^{03}_c \, \mathrm{d}^3 x \right).
\end{align}
Notably, the integrals in the above equation depend solely on the zeroth-order component \(\bar{A}^\mu\) of the wave packet. For the function \(\bar{A}_f^\mu\) given by Eq. \eqref{LG-wave}, we find that \(\int \delta y \bar{T}^{00}_c \, \mathrm{d}^3 x = \int \delta y \bar{T}^{02}_c \, \mathrm{d}^3 x = \int \delta y \bar{T}^{03}_c \, \mathrm{d}^3 x = 0\). The remaining term in Eq. (\ref{expanded_first_term}) can be expressed as:
\begin{equation}\label{rewritten_remaining_term}
	-\frac{2}{E} \tilde{\Gamma}^0_{0 1} \Big|_{\vec{r}_0} \int \delta y \, \bar{T}^{01}_c \, \mathrm{d}^3 x = \frac{1}{E} \tilde{\Gamma}^0_{0 1} \Big|_{\vec{r}_0} \int (\delta x \, \bar{T}^{02}_c - \delta y \, \bar{T}^{01}_c) \, \mathrm{d}^3 x = \frac{1}{E} \tilde{\Gamma}^0_{0 1} \Big|_{\vec{r}_0} \int L^{012} \, \mathrm{d}^3 x,
\end{equation}
due to the axial symmetry in the \(x\)-\(y\) plane, where \(L^{\mu \nu \rho}\) represents the orbital angular momentum tensor.

The spin-dependent part of Eq.~(\ref{analysis5}) can be expressed as
\begin{align}\label{second_part_analysis5}
	&\quad -\frac{2}{E}\tilde{\Gamma}^0_{0 \rho} \Big|_{\vec{r}_0}\int \left(\partial_\lambda (\delta y \, B^{\lambda 0 \rho})- B^{\lambda 0 \rho} \partial_\lambda \delta y \right)\mathrm{d}^3 x \nonumber \\
	&=-\frac{2}{E}\tilde{\Gamma}^0_{0 \rho} \Big|_{\vec{r}_0}\int (\partial_t (\delta y \, B^{0 0 \rho}) \mathrm{d}^3 x + \frac{2}{E} \tilde{\Gamma}^0_{01} \Big|_{\vec{r}_0} \int S^{012} \, \mathrm{d}^3 x,
\end{align}
where the surface integrals have been neglected, and $S^{\lambda\mu\nu}=-S^{\lambda\nu\mu}$ has been considered. Due to the antisymmetry $B^{\lambda\mu\nu}=-B^{\mu\lambda\nu}$, we have $B^{00\rho}=0$. Therefore, the spin-dependent part of Eq.~\eqref{analysis5}  can be simplified to
\begin{equation}\label{third_part_analysis5}
	-\frac{2}{E}\tilde{\Gamma}^0_{0 \rho} \Big|_{\vec{r}_0}\int \left(\partial_\lambda (\delta y \, B^{\lambda 0 \rho})- B^{\lambda 0 \rho} \partial_\lambda \delta y \right)\mathrm{d}^3 x=\frac{2}{E} \tilde{\Gamma}^0_{01} \Big|_{\vec{r}_0} \int S^{012} \, \mathrm{d}^3 x.
\end{equation}

Considering these above results, Eq.~(\ref{analysis5}) can be simplified to:
\begin{equation}\label{final_simplified_equation}
	\frac{d\langle y \rangle_{\sigma,\ell}}{dt}=\frac{1}{E} \tilde{\Gamma}^0_{01} \Big|_{\vec{r}_0} \int(2S^{012}+L^{012}) \mathrm{d}^3 x=\frac{1}{E} \tilde{\Gamma}^0_{01} \Big|_{\vec{r}_0} \left( 2S_z+L_z  \right) \propto (2\sigma + \ell),
\end{equation}
where $L_z=\int L^{012}\mathrm{d}^3 x \propto \ell$ is the intrinsic orbital angular momentum, and $S_z=\int S^{012}\mathrm{d}^3 x\propto \sigma$ represents spin. This relation $d\langle y \rangle_{\sigma,\ell}/dt\propto (2\sigma + \ell)$ suggests that for light with the angular momentum where $\sigma=\ell$, the transverse deviation induced by its spin should be twice that induced by its intrinsic orbital angular momentum.

\section{Transverse deviations in realistic optical materials}\label{The observation of the OHE in two realistic optical materials}

The materials selected for practical application are derived from the articles \cite{sheng2013trapping,sheng2016wavefront}, which were originally employed to investigate the gravitational lensing effects predicted by Einstein's theory. The careful selection of materials is crucial, as it may enhance our ability to observe deflection effects more clearly compared to the subtle effects encountered in the actual universe. Furthermore, it is beneficial to revisit the distinctions between the gravitational OHE and the gravitational SHE.

\subsection{The deviation angle \(\theta_{\sigma,\ell}\) in the first material}

\begin{figure*}   
	\centering 
	\includegraphics[width=0.7\textwidth]{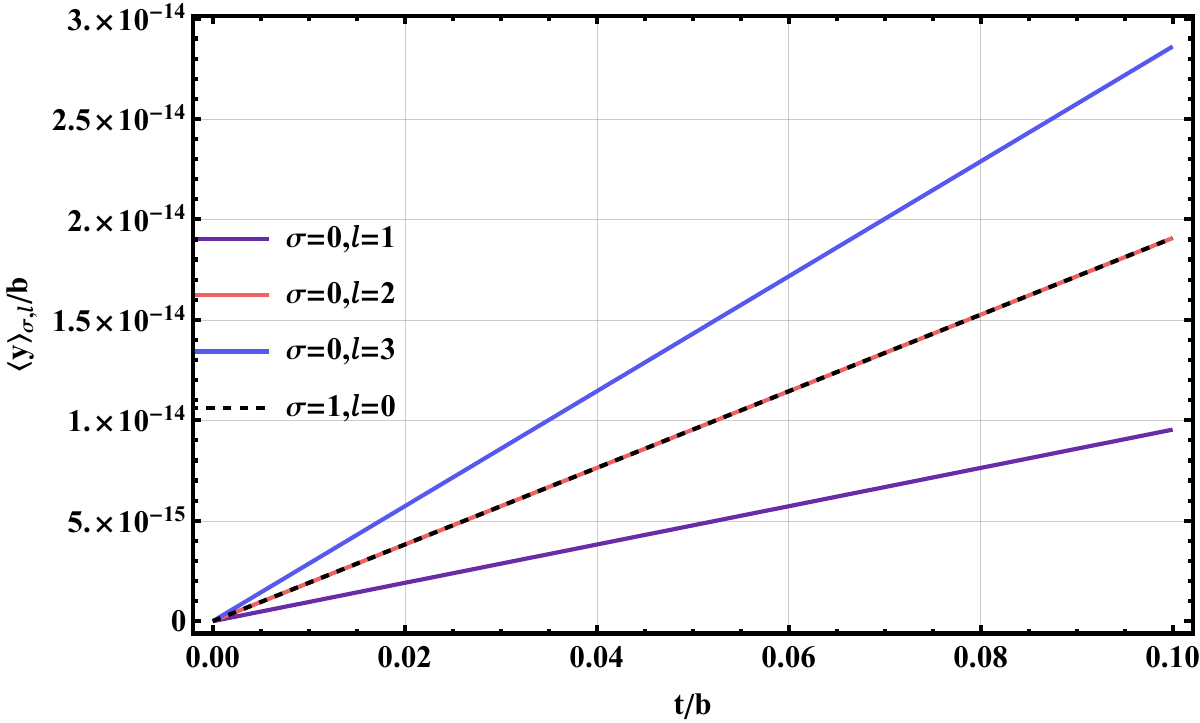}
	\caption{The transverse trajectory of the wave packet with $\sigma$ and $\ell$ in the first optical material. Here $b=10^2$, $\lambda=6.05\times10^{-9}b$, $\sigma_{\perp}=\sigma_{z}=5\times10^{-3}b$, and all parameters are nondimensionalized. }\label{fig.res1}
\end{figure*}

The refractive index for the first material is given by $n_1 = n_0 (1 + (a/r)^4)^{1/2}$, where \(n_0 = 1.0488\), \(a = 28.5 \, \micro\meter\), and \(r = \sqrt{x^2 + y^2 + z^2}\) \cite{sheng2013trapping}. Consequently, the metric is:
\begin{equation}\label{pert-h}
	\eta'_{\mu\nu} = \begin{pmatrix}
		-n_0^{-2} & 0 & 0 & 0 \\
		0 & 1 & 0 & 0 \\
		0 & 0 & 1 & 0 \\
		0 & 0 & 0 & 1 \\
	\end{pmatrix}, \quad
	h_{\mu\nu} = n_0^{-2} \left(\frac{a}{r}\right)^4 \begin{pmatrix}
		1 & 0 & 0 & 0 \\
		0 & 0 & 0 & 0 \\
		0 & 0 & 0 & 0 \\
		0 & 0 & 0 & 0 \\
	\end{pmatrix}.
\end{equation}
The simplified equations of motion can be found in Appendix \ref{Simplifications of the equations of motion for the first material}. Through numerical calculations, the transverse trajectory of an electromagnetic wave packet with spin \(\sigma\) and intrinsic orbital angular momentum \(\ell\) along the \(y\)-axis is illustrated in Fig. \ref{fig.res1}. It is evident that the factor \(\langle y\rangle_{\sigma,0}/\sigma\) is twice that of \(\langle y\rangle_{0,\ell}/\ell\). Consequently, the transverse displacement \(\langle y\rangle_{\sigma,\ell}\) is expected to be proportional to \(2\sigma + \ell\).

Within the regime \(t \ll b\), the transverse trajectory of this wave packet can be approximated as:
\begin{equation}\label{transverse trajectory 1}
	\langle y \rangle_{\sigma,\ell} \simeq \frac{(2\sigma + \ell) a^4 \lambda t}{2\pi n_0 b^5},
\end{equation}
where \(\lambda = 2\pi n_0/p_z\). Therefore, the transverse velocity is then given by:
\begin{equation}
	\frac{d\langle y \rangle_{\sigma,\ell}}{dt} \simeq \frac{(2\sigma + \ell) a^4 \lambda}{2\pi n_0 b^5}.
\end{equation}
As this wave packet moves away from its initial position, predominantly along the $z$-axis, its distance from the center of the material can be approximately described by \(\sqrt{b^2 + t^2}\). Consequently, the wave packet's transverse velocity can be approximated as:
\begin{equation}
	\frac{d\langle y \rangle_{\sigma,\ell}}{dt} \simeq \frac{(2\sigma + \ell) a^4 \lambda}{2\pi n_0} \frac{1}{(t^2 + b^2)^{\frac{5}{2}}}.
\end{equation}

Unlike the case in Lense-Thirring spacetime, the size of the material is not infinite and has finite boundaries. Therefore, we must consider a finite moving distance for the wave packet (see Figure \ref{fig:picture}). Assume it moves from \(z = -d\) to \(z = +z_0\) and finally exits the boundary into the detector. Considering the initial conditions: when \(t = 0\), \(y = 0\), and \(d\langle y \rangle_{\sigma,\ell}/dt = 0\), the transverse trajectory becomes:
\begin{equation}
	\begin{split}
		\langle y \rangle_{\sigma,\ell} &\simeq \frac{(2\sigma + \ell) a^4 \lambda}{2\pi n_0} \left( -\frac{(d - t)/b}{\sqrt{((d - t)/b)^2 + 1}} + \frac{1}{3} \frac{((d - t)/b)^3}{(((d - t)/b)^2 + 1)^{\frac{3}{2}}} + \frac{(d/b)}{\sqrt{(d/b)^2 + 1}} - \right. \\
		& \quad \left. \frac{1}{3} \frac{(d/b)^3}{((d/b)^2 + 1)^{\frac{3}{2}}} \right) - \frac{(2\sigma + \ell) a^4 \lambda}{2\pi n_0 (b^2 + d^2)^{\frac{5}{2}}} t.
	\end{split}
\end{equation}

Similar to curved spacetime, we can also define the \(x-z\) plane as the null geodesic plane within the material. Therefore, within \(\Delta t \simeq (d + z_0)/n_0\), the maximum deflection angle perpendicular to the null geodesic plane is:
\begin{equation}
	\theta_{\sigma,\ell} \simeq \frac{(2\sigma + \ell) a^4 \lambda}{2\pi n_0} \left( \frac{1}{\left(\frac{d + z_0}{n_0} - d\right)^2 + b^2)^{\frac{5}{2}}} - \frac{1}{(b^2 + d^2)^{\frac{5}{2}}} \right).
\end{equation}
By referencing the specifications and sizes of materials achievable in the laboratory as described in the literature \cite{sheng2013trapping}, we can take \(d = 100.3 \, \micro\meter\), \(b = 43.24 \, \micro\meter\), \(z_0 = 109.6 \, \micro\meter\), yielding:
\begin{equation}\label{theta1}
	\theta_{\sigma,\ell} \simeq 1.28 \times 10^{-7}(2\sigma+\ell)\lambda ~ \text{rad},
\end{equation}
where the wavelength $\lambda$ is in micrometers.

\subsection{The Deviation Angle \(\theta_{\sigma,\ell}\) in the Second Material}

\begin{figure*}   
	\centering 
    \includegraphics[width=0.7\textwidth]{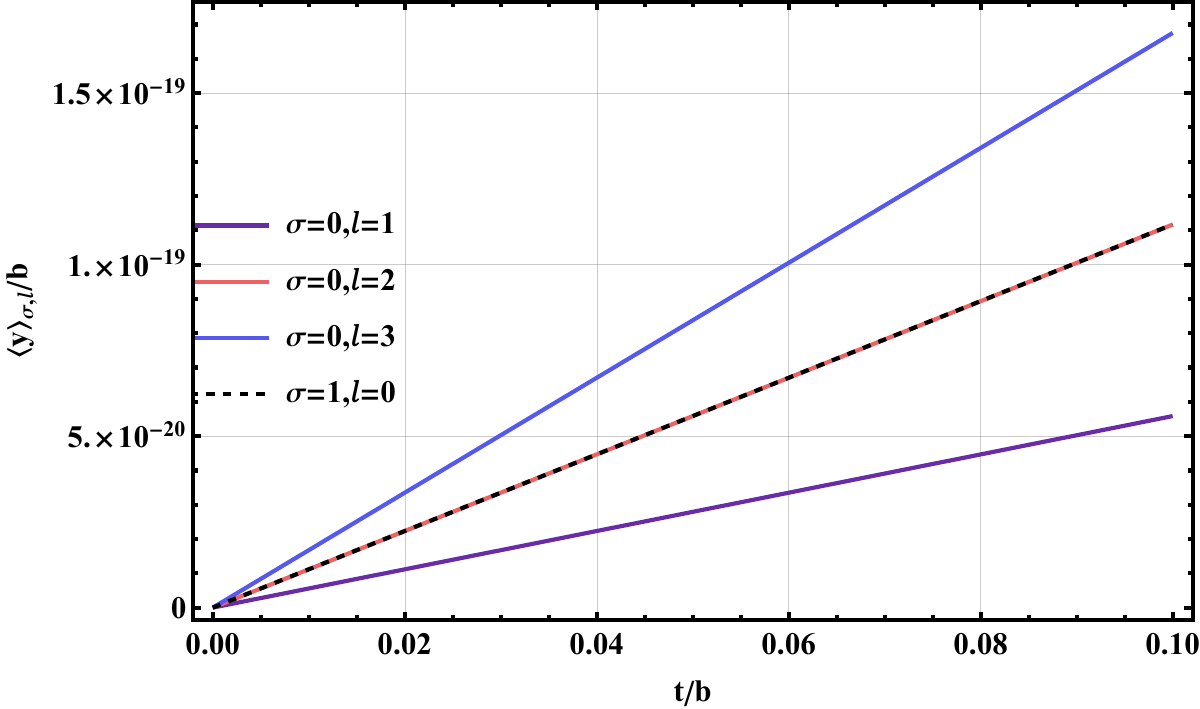}
	\caption{The transverse trajectory of the wave packet with $\sigma$ and $\ell$ in the second optical material. Here $b=10^2$, $\lambda=6.15\times10^{-9}b$, $\sigma_{\perp}=\sigma_{z}=5\times10^{-3}b$, and all parameters are nondimensionalized. }\label{fig.res2}
\end{figure*}

The refractive index for the second material is given by $n_2 = n_0 + a/(1 + \frac{(x^2 + z^2)^4}{r_c^8})$, where \(r_c = 9.69 \, \mu\text{m}\), \(a = 0.0922\) (dimensionless), and \(n_0 = 1.37\) \cite{sheng2016wavefront}. Consequently, the metric is:
\begin{equation}\label{pert-h2}
	\eta'_{\mu\nu} = \begin{pmatrix}
		-n_0^{-2} & 0 & 0 & 0 \\
		0 & 1 & 0 & 0 \\
		0 & 0 & 1 & 0 \\
		0 & 0 & 0 & 1 \\
	\end{pmatrix}, \quad
	h_{\mu\nu} = \frac{2a}{n_0} \left(1 + \frac{(x^2 + z^2)^4}{r_c^8}\right)^{-1} \begin{pmatrix}
		1 & 0 & 0 & 0 \\
		0 & 0 & 0 & 0 \\
		0 & 0 & 0 & 0 \\
		0 & 0 & 0 & 0 \\
	\end{pmatrix}.
\end{equation}
The simplified equations of motion can be found in Appendix \ref{Simplifications of the equations of motion for the second material}. The transverse trajectory of the wave packet in the second material is shown in Fig. \ref{fig.res2}. Similar to the first material, the transverse displacement \(\langle y\rangle_{\sigma,\ell}\) is expected to be proportional to \(2\sigma + \ell\).

Within the regime \(t \ll b\), the transverse trajectory for the wave packet with $\sigma$ and $\ell$ can be approximated as:
\begin{equation}\label{transverse trajectory 2}
	\langle y \rangle_{\sigma,\ell} \simeq \frac{(2\sigma + \ell) a \lambda r_c^8t}{2\pi n_0 b^9},
\end{equation}
where \(\lambda = 2\pi n_0/p_z\) and the transverse velocity is:
\begin{equation}
	\frac{d\langle y \rangle_{\sigma,\ell}}{dt} \simeq \frac{(2\sigma + \ell) a \lambda r_c^8 }{2\pi n_0 (b^2 + t^2)^{\frac{9}{2}}}.
\end{equation}
When considering this wave packet moving from \(z = -d\) to \(z = +z_0\), the transverse trajectory can be given by:
\begin{equation}
	\begin{split}
		\langle y\rangle_{\sigma,\ell} &\simeq \frac{(2\sigma + \ell) a \lambda r_c^8}{2\pi n_0 b^8} \left( -\frac{(d - t)/b}{\sqrt{((d - t)/b)^2 + 1}} + \frac{d/b}{\sqrt{(d/b)^2 + 1}} + \frac{((d - t)/b)^3}{(((d - t)/b)^2 + 1)^{\frac{3}{2}}} - \right. \\
		& \quad \left. \frac{(d/b)^3}{((d/b)^2 + 1)^{\frac{3}{2}}} + \frac{3}{5} \frac{((d - t)/b)^5}{(((d - t)/b)^2 + 1)^{\frac{5}{2}}} - \frac{3}{5} \frac{(d/b)^5}{((d/b)^2 + 1)^{\frac{5}{2}}} - \right. \\
		& \quad \left. \frac{1}{7} \frac{((d - t)/b)^7}{(((d - t)/b)^2 + 1)^{\frac{7}{2}}} + \frac{1}{7} \frac{(d/b)^7}{((d/b)^2 + 1)^{\frac{7}{2}}} \right) - \frac{(2\sigma + \ell) a \lambda r_c^8 t}{2\pi n_0 (b^2 + d^2)^{\frac{9}{2}}}.
	\end{split}
\end{equation}
As \(\Delta t \simeq (d + z_0)/n_0\), the deviation angle from the null geodesic plane is:
\begin{equation}
	\theta_{\sigma,\ell} \simeq \frac{(2\sigma + \ell) a r_c^8 \lambda}{2\pi n_0} \left( \frac{1}{\left(\frac{d + z_0}{n_0} - d\right)^2 + b^2)^{\frac{9}{2}}} - \frac{1}{(b^2 + d^2)^{\frac{9}{2}}} \right).
\end{equation}

Referring to the specifications and dimensions of materials achievable in the laboratory as described in the literature \cite{sheng2016wavefront}, we consider \(d = 42.8 \, \micro\meter\), \(b = 39.28 \, \micro\meter\), \(z_0 = 200 \, \micro\meter\), leading to:
\begin{equation}\label{theta2}
	\theta_{\sigma,\ell} \simeq 1.10 \times 10^{-10}(2\sigma+\ell) \lambda~ \text{rad},
\end{equation}
where the wavelength $\lambda$ is in micrometers. Compared to Eq. \eqref{theta1}, this angle $\theta_{\sigma,\ell}$ is significantly smaller. This suggests that the first optical material may be more suitable for observing the spin and orbital Hall effects than the second material.

\section{Discussion}\label{Discussion}

In this study, we examine the spin Hall effects and orbital Hall effects of light as it propagates through inhomogeneous optical materials. When an electromagnetic wave packet with spin \(\sigma\) and intrinsic orbital angular momentum \(\ell\) travels through such a medium, its trajectory deviates from that of a wave packet without angular momentum ($\sigma=0$ and $\ell=0$) by an angle \(\theta_{\sigma,\ell} \propto (2\sigma+\ell)\). Notably, the spin \(\sigma\) contributes a factor of 2, while  the intrinsic orbital angular momentum \(\ell\) contributes a factor of 1. This indicates that spin has a greater influence on the angle \(\theta_{\sigma,\ell}\) than intrinsic orbital angular momentum, highlighting  the different contributions of these two types of angular momentum  in light propagation.

For the optical materials considered in this work, the angle $\theta_{\sigma,\ell}$ is much smaller than $1\,\radian$,  allowing the transverse displacement to be approximated as $\delta y_{\sigma,\ell} \sim \theta_{\sigma,\ell} D $. Here \(D\) denotes the propagation distance. This displacement \(\delta y_{\sigma,\ell}\), associated with intrinsic angular momentum $\sigma$ and $\ell$, has been experimentally observed \cite{wang2021generalized, hosten2008observation, dasgupta2005experimental}. In these experiments, \(\delta y_{\sigma,\ell}\) can be as small as the order of the wavelength of light, \(\lambda\).

In the first material examined, the propagation distance is limited to \(D \simeq 109.6 \, \micro\meter\). According to Eq. \eqref{theta1}, as the wave packet passes through the material, the angle is approximated as \(\theta_{\sigma,\ell} \simeq 1.28 \times 10^{-7}(2\sigma + \ell) \lambda\, \radian\), where the wavelength \(\lambda\) is in micrometers. Thus, the transverse displacement is approximated as \(\delta y_{\sigma,\ell} \sim 1.40 \times 10^{-5}(2\sigma + \ell) \lambda\), which is too small to be detected. However, the wave packet can continue to propagate after leaving the material, with the angle \(\theta_{\sigma,\ell}\) remaining constant. As the propagation distance increases, so does the transverse displacement. When \(D\) reaches approximately \(1\,\meter\), the transverse displacement becomes \(\delta y_{\sigma,\ell} \sim 0.128 (2\sigma + \ell) \lambda\). In this case, the transverse displacement \(\delta y_{\sigma,\ell}\) is on the order of the wavelength \(\lambda\), making it comparable to the wavelength \(\lambda\) and potentially observable in experiments. 

Specifically, for polarized light with \(\sigma=1\) and \(\ell=0\), the displacement for the SHE is approximately \(\delta y_{1,0} \sim 0.256\lambda\). In contrast, for vortex light, the orbital angular momentum \(\ell\) can reach magnitudes on the order of \(10^4\) \cite{courtial1997gaussian}, resulting in a displacement \(\delta y_{0,10^4} \sim1.28 \times 10^3 \lambda\). This indicates that, in experimental settings, the OHE is more readily observable than the SHE.

\section*{Acknowledgement}\label{Acknowledgement}
 This work is supported by National Key R\&D Program of China (Grant No.2024YFE0109802), the National Natural Science Foundation of China (Grant No. 12375084 and No. 12247101), and the Fundamental Research Funds for the Central Universities (Grant No. lzujbky-2024-jdzx06).

\appendix
\section{Simplifications of the equations of motion for the first material}\label{Simplifications of the equations of motion for the first material}
In the first material, the metric is:
\begin{equation}
	\eta'_{\mu\nu}=\begin{pmatrix}
		-n^{-2}_0&0&0&0\\
		0&1&0&0\\
		0&0&1&0\\
		0&0&0&1\\
	\end{pmatrix},\quad
	h_{\mu\nu}=n_0^{-2}(\frac{a}{r})^4\begin{pmatrix}
		1&0&0&0\\
		0&0&0&0\\
		0&0&0&0\\
		0&0&0&0\\
	\end{pmatrix}.
\end{equation}
Accordingly, the nonzero first order Christoffel symbols: 
\begin{equation}
	\begin{split}
		&\tilde{\Gamma}^t_{tx}=\tilde{\Gamma}^t_{xt}=\frac{2a^4x}{r^6}, \quad \tilde{\Gamma}^t_{ty}=\tilde{\Gamma}^t_{yt}=\frac{2a^4y}{r^6}, \quad
	\tilde{\Gamma}^t_{tz}=\tilde{\Gamma}^t_{zt}=\frac{2a^4z}{r^6}, \\
		&\tilde{\Gamma}^x_{tt}=\frac{2a^4x}{n_0^2 r^6},\quad
		\tilde{\Gamma}^y_{tt}=\frac{2a^4y}{n_0^2 r^6}, \quad
		\tilde{\Gamma}^z_{tt}=\frac{2a^4z}{n_0^2 r^6}. \quad
	\end{split}
\end{equation}
The nonzero first order Ricci tensors: 
\begin{equation}
	\begin{split}
	&\tilde{R}_{xx}=-\frac{2a^4(y^2+z^2-5x^2)}{r^8},\quad
	\tilde{R}_{yy}=-\frac{2a^4(x^2+z^2-5y^2)}{r^8},\quad
	\tilde{R}_{zz}=-\frac{2a^4(x^2+y^2-5z^2)}{r^8},\\
		&\tilde{R}_{tt}=-\frac{6a^4}{n_0^2r^6},\quad
		\tilde{R}_{xy}=\tilde{R}_{yx}=\frac{12a^4xy}{r^8},\quad
		\tilde{R}_{xz}=\tilde{R}_{zx}=\frac{12a^4xz}{r^8},\quad
		\tilde{R}_{yz}=\tilde{R}_{zy}=\frac{12a^4yz}{r^8}.
	\end{split}
\end{equation}
The Eqs.\eqref{final equations} becomes:
\begin{equation}
	\begin{array}{l}
		(-n_0^2\partial_t^2+\partial_{i}^2)\tilde{A}^{0}=\frac{4n_0^2a^4}{r^6}(x\partial_{t}\bar{A}^1+y\partial_{t}\bar{A}^2+z\partial_{t}\bar{A}^3)\\
		(-n_0^2\partial_t^2+\partial_{i}^2)\tilde{A}^1=\frac{n_0^2a^4}{r^4}(\partial_{t}^2\bar{A}^1)-\frac{2a^4}{r^6}(x\partial_{x}\bar{A}^1+y\partial_{y}\bar{A}^1+z\partial_{z}\bar{A}^1)-\frac{2a^4}{r^8}(6x(y\bar{A}^2+z\bar{A}^3)\\+(5x^2-y^2-z^2)\bar{A}^1)\\
		(-n_0^2\partial_t^2+\partial_{i}^2)\tilde{A}^2=\frac{n_0^2a^4}{r^4}(\partial_{t}^2\bar{A}^2)-\frac{2a^4}{r^6}(x\partial_{x}\bar{A}^2+y\partial_{y}\bar{A}^2+z\partial_{z}\bar{A}^2)-\frac{2a^4}{r^8}(6y(x\bar{A}^1+z\bar{A}^3)\\+(5y^2-x^2-z^2)\bar{A}^2)\\
		(-n_0^2\partial_t^2+\partial_{i}^2)\tilde{A}^3=\frac{n_0^2a^4}{r^4}(\partial_{t}^2\bar{A}^3)-\frac{2a^4}{r^6}(x\partial_{x}\bar{A}^3+y\partial_{y}\bar{A}^3+z\partial_{z}\bar{A}^3)-\frac{2a^4}{r^8}(6z(x\bar{A}^1+y\bar{A}^2)\\+(5z^2-x^2-y^2)\bar{A}^3).\\
	\end{array}
\end{equation}
Expand them at $(-b,0,0)$:
\begin{equation}
	\begin{array}{l}
		(-n_0^2\partial_t^2+\partial_{i}^2)\tilde{A}^{0}\simeq\frac{4n_0^2a^4}{b^6}((1-\frac{5x}{b})\partial_{t}\bar{A}^1+\frac{y}{b}\partial_{t}\bar{A}^2+\frac{z}{b}\partial_{t}\bar{A}^3)\\
		
		(-n_0^2\partial_t^2+\partial_{i}^2)\tilde{A}^1\simeq\frac{n_0^2a^4}{b^4}((1-\frac{4x}{b})\partial_{t}^2\bar{A}^1)-\frac{2a^4}{b^5}((1-\frac{5x}{b})\partial_{x}\bar{A}^1+\frac{y}{b}\partial_{y}\bar{A}^1+\frac{z}{b}\partial_{z}\bar{A}^1)-\frac{2a^4}{b^6}(\frac{6y}{b}\bar{A}^2\\+\frac{6z}{b}\bar{A}^3)+5(1-\frac{6x}{b})\bar{A}^1)\\
		
		(-n_0^2\partial_t^2+\partial_{i}^2)\tilde{A}^2\simeq\frac{n_0^2a^4}{b^4}((1-\frac{4x}{b})\partial_{t}^2\bar{A}^2)-\frac{2a^4}{b^5}((1-\frac{5x}{b})\partial_{x}\bar{A}^2+\frac{y}{b}\partial_{y}\bar{A}^2+\frac{z}{b}\partial_{z}\bar{A}^2)-\frac{2a^4}{b^6}(\frac{6y}{b}\bar{A}^1\\-(1-\frac{6x}{b})\bar{A}^2)\\
		
		(-n_0^2\partial_t^2+\partial_{i}^2)\tilde{A}^3\simeq\frac{n_0^2a^4}{b^4}((1-\frac{4x}{b})\partial_{t}^2\bar{A}^3)-\frac{2a^4}{b^5}((1-\frac{5x}{b})\partial_{x}\bar{A}^3+\frac{y}{b}\partial_{y}\bar{A}^3+\frac{z}{b}\partial_{z}\bar{A}^3)-\frac{2a^4}{b^6}(\frac{6z}{b}\bar{A}^1\\-(1-\frac{6x}{b})\bar{A}^3).\\
	\end{array}
\end{equation}
These equations actually constitute a system of second-order non-homogeneous partial differential equations with variable coefficients. We consider applying Fourier transform to the zeroth-order wave function $\bar{A}^i$:
\begin{equation}
	\bar{A}^\rho_f=\frac{1}{2\pi^3}\int\bar{A}^\rho\exp(-i\vec{k}\cdot\vec{x})\text{d}^3x, \quad \widetilde{A}^\rho_f=\frac{1}{2\pi^3}\int\widetilde{A}^\rho\exp(-i\vec{k}\cdot\vec{x})\text{d}^3x. \quad
\end{equation} 
We can then make some substitutions to some of the terms:
\begin{equation}
	\partial_{x^i} \bar{A}^{j} \rightarrow i k_{x^i} \bar{A}_f^{j},\quad \partial_{t}  \bar{A}^{j} \rightarrow -i \omega \bar{A}_f^{j},\quad x^i \bar{A}^{j} \rightarrow i\partial_{k_{x^i}} \bar{A}_f^{j}.
\end{equation}
After some simplifications, we finally yield the equations of motion in momentum space:
\begin{equation}
	\begin{array}{l}
		(\partial_t^2+\omega^2)\tilde{A}_f^{0}=-\frac{4a^4}{b^5}((1-\frac{5i}{b}\partial_{k_x})(-i\omega\bar{A}_f^1)+\frac{i}{b}\partial_{k_y}(-i\omega\bar{A}_f^2)+\frac{i}{b}\partial_{k_z}(-i\omega\bar{A}_f^3))\\
		
		(\partial_t^2+\omega^2)\tilde{A}_f^{1}=\frac{a^4}{b^4}(1-\frac{4i}{b}\partial_{k_x})(\omega^2\bar{A}_f^1)+\frac{2a^4}{n_0^2b^5}((1-\frac{5i}{b}\partial_{k_x})(ik_x\bar{A}_f^1)+\frac{i}{b}\partial_{k_y}(ik_y\bar{A}_f^1)+\frac{i}{b}\partial_{k_z}(ik_z\bar{A}_f^1))\\+\frac{2a^4}{n_0^2b^6}(\frac{6i}{b}\partial_{k_y}\bar{A}^2_f+\frac{6i}{b}\partial_{k_z}\bar{A}^3_f+5(1-\frac{6i}{b}\partial_{k_x})\bar{A}^1_f)\\
		
		(\partial_t^2+\omega^2)\tilde{A}_f^{2}=\frac{a^4}{b^4}(1-\frac{4i}{b}\partial_{k_x})(\omega^2\bar{A}_f^2)+\frac{2a^4}{n_0^2b^5}((1-\frac{5i}{b}\partial_{k_x})(ik_x\bar{A}_f^2)+\frac{i}{b}\partial_{k_y}(ik_y\bar{A}_f^2)+\frac{i}{b}\partial_{k_z}(ik_z\bar{A}_f^2))\\+\frac{2a^4}{n_0^2b^6}(\frac{6i}{b}\partial_{k_y}\bar{A}^1_f-(1-\frac{6i}{b}\partial_{k_x})\bar{A}^2_f)\\
		
		(\partial_t^2+\omega^2)\tilde{A}_f^{3}=\frac{a^4}{b^4}(1-\frac{4i}{b}\partial_{k_x})(\omega^2\bar{A}_f^3)+\frac{2a^4}{n_0^2b^5}((1-\frac{5i}{b}\partial_{k_x})(ik_x\bar{A}_f^3)+\frac{i}{b}\partial_{k_y}(ik_y\bar{A}_f^3)+\frac{i}{b}\partial_{k_z}(ik_z\bar{A}_f^3))\\+\frac{2a^4}{n_0^2b^6}(\frac{6i}{b}\partial_{k_z}\bar{A}^1_f-(1-\frac{6i}{b}\partial_{k_x})\bar{A}^3_f)\\
	\end{array}
\end{equation}
\section{Simplifications of the equations of motion for the second material}\label{Simplifications of the equations of motion for the second material}
The metric in the second material is:
\begin{equation}
	\eta'_{\mu\nu}=\begin{pmatrix}
		-n^{-2}_0&0&0&0\\
		0&1&0&0\\
		0&0&1&0\\
		0&0&0&1\\
	\end{pmatrix},\quad
	h_{\mu\nu}=\frac{2a}{n_0}(1+\frac{(x^2+z^2)^4}{r_c^8})^{-1}\begin{pmatrix}
		1&0&0&0\\
		0&0&0&0\\
		0&0&0&0\\
		0&0&0&0\\
	\end{pmatrix}.
\end{equation}
Correspondingly, the nonzero first order Christoffel symbols: 
\begin{equation}
	\begin{split}
		\tilde{\Gamma}^t_{tx}&=\tilde{\Gamma}^t_{xt}=\frac{8n_0ar_c^8x(x^2+z^2)^3}{(r_c^8+(x^2+z^2)^4)^2}, \quad
		\tilde{\Gamma}^t_{tz}=\tilde{\Gamma}^t_{zt}=\frac{8n_0ar_c^8z(x^2+z^2)^3}{(r_c^8+(x^2+z^2)^4)^2}, \\
		\tilde{\Gamma}^x_{tt}&=\frac{8ar_c^8x(x^2+z^2)^3}{n_0(r_c^8+(x^2+z^2)^4)^2},\quad
		\tilde{\Gamma}^z_{tt}=\frac{8ar_c^8z(x^2+z^2)^3}{n_0(r_c^8+(x^2+z^2)^4)^2}. \quad
	\end{split}
\end{equation}
Nonzero first order Ricci tensors: 
\begin{equation}
	\begin{split}
		\tilde{R}_{xx}&=-\frac{8n_0ar_c^8(-(9x^2-z^2)(x^2+z^2)^6+r_c^8((x^2+z^2)^2(7x^2+z^2)))}{(r_c^8+(x^2+z^2)^4)^3},\\
		\tilde{R}_{zz}&=-\frac{8n_0ar_c^8((x^2-9z^2)(x^2+z^2)^6+r_c^8((x^2+z^2)^2(x^2+7z^2)))}{(r_c^8+(x^2+z^2)^4)^3},\\
		\tilde{R}_{tt}&=\frac{64ar_c^8(r_c^8(x^2+z^2)^3-(x^2+z^2)^7)}{n_0(r_c^8+(x^2+z^2)^4)^3},\\
		\tilde{R}_{xz}&=\tilde{R}_{zx}=\frac{16n_0ar_c^8xz(x^2+z^2)^2(-3r_c^8+5(x^2+z^2)^4))}{(r_c^8+(x^2+z^2)^4)^3}.\quad	
	\end{split}
\end{equation}
The Eqs.\eqref{final equations} becomes:
\begin{equation}
	\begin{array}{l}
		(-n_0^2\partial_t^2+\partial_{i}^2)\tilde{A}^{0}=\frac{16n_0^3ar_c^8}{(r_c^8+(x^2+z^2)^4)^2}(x^2+z^2)^3(x\partial_{t}\bar{A}^1+z\partial_{t}\bar{A}^3)\\
		
		(-n_0^2\partial_t^2+\partial_{i}^2)\tilde{A}^1=\frac{2ar_c^8}{(r_c^8+(x^2+z^2)^4)^3}((n_0^3(r_c^8+(x^2+z^2)^4)^2\partial_{t}^2\bar{A}^1)-4n_0(r_c^8+(x^2+z^2)^4)(x^2+z^2)^3\\(x\partial_{x}\bar{A}^1+z\partial_{z}\bar{A}^1)+4n_0(x^2+z^2)^2(-(9x^2-z^2)(x^2+z^2)^6+r_c^8((x^2+z^2)^2(7x^2+z^2)))\bar{A}^1-\\8n_0xz(x^2+z^2)^2(-3r_c^8+5(y^2+z^2)^4)\bar{A}^3)\\
		
		(-n_0^2\partial_t^2+\partial_{i}^2)\tilde{A}^2=\frac{2an_0r_c^8}{(r_c^8+(x^2+z^2)^4)^2}((n_0^2(r_c^8+(x^2+z^2)^4)^2\partial_{t}^2\bar{A}^2)-4(x^2+z^2)^3(x\partial_{x}\bar{A}^2+z\partial_{z}\bar{A}^2))\\
		
		(-n_0^2\partial_t^2+\partial_{i}^2)\tilde{A}^3=\frac{2ar_c^8}{(r_c^8+(x^2+z^2)^4)^3}((n_0^3(r_c^8+(x^2+z^2)^4)^2\partial_{t}^2\bar{A}^3)-4n_0(r_c^8+(x^2+z^2)^4)(x^2+z^2)^3\\(x\partial_{x}\bar{A}^3+z\partial_{z}\bar{A}^3)+4n_0(x^2+z^2)^2((x^2-9z^2)(x^2+z^2)^6+r_c^8((x^2+z^2)^2(x^2+7z^2)))\bar{A}^3-\\8n_0xz(x^2+z^2)^2(-3r_c^8+5(y^2+z^2)^4)\bar{A}^1).\\
	\end{array}
\end{equation}
Similar to the following procedures, we can rewrite the equations in momentum space:
\begin{equation}
	\begin{array}{l}
		(\partial_t^2+\omega^2)\tilde{A}_f^{0}=-\frac{2}{n_0^2}(\hat{P}_{xf}(-i\omega\bar{A}^1_f)+\hat{P}_{zf}(-i\omega\bar{A}^3_f))\\

		(\partial_t^2+\omega^2)\tilde{A}_f^{1}=\frac{1}{n_0^2}\hat{G}_f(\omega^2\bar{A}^1_f)+\frac{1}{n_0^4}(\hat{P}_{xf}(ik_x\bar{A}^1_f)+\hat{P}_{zf}(ik_z\bar{A}^1_f))-\frac{1}{n_0^2}\hat{O}_f\bar{A}^1_f+\frac{1}{n_0^2}\hat{Y}_f\bar{A}^3_f\\
		
		(\partial_t^2+\omega^2)\tilde{A}_f^{2}=\frac{1}{n_0^2}\hat{G}_f(\omega^2\bar{A}^2_f)+\frac{1}{n_0^4}(\hat{P}_{xf}(ik_x\bar{A}^2_f)+\hat{P}_{zf}(ik_z\bar{A}^2_f))\\
		
		(\partial_t^2+\omega^2)\tilde{A}_f^{3}=\frac{1}{n_0^2}\hat{G}_f(\omega^2\bar{A}^3_f)+\frac{1}{n_0^4}(\hat{P}_{xf}(ik_x\bar{A}^3_f)+\hat{P}_{zf}(ik_z\bar{A}^3_f))-\frac{1}{n_0^2}\hat{B}_f\bar{A}^3_f+\frac{1}{n_0^2}\hat{Y}_f\bar{A}^1_f.\\
	\end{array}
\end{equation}
where the operators:
\begin{equation}
\begin{split}
	&\hat{G}_f=\frac{2an_0^3r_c^8}{b^8+r_c^8}-\frac{16ab^7n_0^3r_c^8}{(b^8+r_c^8)^2}(i\partial_{k_x}),\quad
	\hat{P}_{zf}=\frac{8ab^6n_0^3r_c^8}{(b^8+r_c^8)^2}(i\partial_{k_z}),\\
	&\hat{Y}_{f}=\frac{16ab^5n_0r_c^8(5b^8-3r_c^8)}{(b^8+r_c^8)^3}(i\partial_{k_z}),\\
	&\hat{P}_{xf}=\frac{8ab^7n_0^3r_c^8}{(b^8+r_c^8)^2}+\frac{8ab^7n_0^3r_c^8(-9b^{14}+7b^6r_c^8)}{(b^8+r_c^8)^3}(i\partial_{k_x}),\\
	&\hat{B}_{f}=\frac{8ab^6n_0r_c^8}{(b^8+r_c^8)^2}-\frac{16ab^5n_0r_c^8(5b^8-3r_c^8)}{(b^8+r_c^8)^3}(i\partial_{k_x}),\\
	&\hat{O}_{f}=-\frac{8ab^6n_0r_c^8(9b^{8}-7r_c^8)}{(b^8+r_c^8)^3}+\frac{48n_0ar_c^8(15b^{21}-42b^{13}r_c^8+7b^5r_c^16)}{(b^8+r_c^8)^4}(i\partial_{k_x}).
\end{split}
\end{equation}

\section*{References}
\bibliography{reference2.bib} 

\begin{thebibliography}{46}%
\makeatletter
\providecommand \@ifxundefined [1]{%
 \@ifx{#1\undefined}
}%
\providecommand \@ifnum [1]{%
 \ifnum #1\expandafter \@firstoftwo
 \else \expandafter \@secondoftwo
 \fi
}%
\providecommand \@ifx [1]{%
 \ifx #1\expandafter \@firstoftwo
 \else \expandafter \@secondoftwo
 \fi
}%
\providecommand \natexlab [1]{#1}%
\providecommand \enquote  [1]{``#1''}%
\providecommand \bibnamefont  [1]{#1}%
\providecommand \bibfnamefont [1]{#1}%
\providecommand \citenamefont [1]{#1}%
\providecommand \href@noop [0]{\@secondoftwo}%
\providecommand \href [0]{\begingroup \@sanitize@url \@href}%
\providecommand \@href[1]{\@@startlink{#1}\@@href}%
\providecommand \@@href[1]{\endgroup#1\@@endlink}%
\providecommand \@sanitize@url [0]{\catcode `\\12\catcode `\$12\catcode `\&12\catcode `\#12\catcode `\^12\catcode `\_12\catcode `\%12\relax}%
\providecommand \@@startlink[1]{}%
\providecommand \@@endlink[0]{}%
\providecommand \url  [0]{\begingroup\@sanitize@url \@url }%
\providecommand \@url [1]{\endgroup\@href {#1}{\urlprefix }}%
\providecommand \urlprefix  [0]{URL }%
\providecommand \Eprint [0]{\href }%
\providecommand \doibase [0]{https://doi.org/}%
\providecommand \selectlanguage [0]{\@gobble}%
\providecommand \bibinfo  [0]{\@secondoftwo}%
\providecommand \bibfield  [0]{\@secondoftwo}%
\providecommand \translation [1]{[#1]}%
\providecommand \BibitemOpen [0]{}%
\providecommand \bibitemStop [0]{}%
\providecommand \bibitemNoStop [0]{.\EOS\space}%
\providecommand \EOS [0]{\spacefactor3000\relax}%
\providecommand \BibitemShut  [1]{\csname bibitem#1\endcsname}%
\let\auto@bib@innerbib\@empty
\bibitem [{\citenamefont {Bliokh}\ and\ \citenamefont {Desyatnikov}(2009)}]{bliokh2009spin}%
  \BibitemOpen
  \bibfield  {author} {\bibinfo {author} {\bibfnamefont {K.~Y.}\ \bibnamefont {Bliokh}}\ and\ \bibinfo {author} {\bibfnamefont {A.~S.}\ \bibnamefont {Desyatnikov}},\ }\href@noop {} {\bibfield  {journal} {\bibinfo  {journal} {Physical Review A}\ }\textbf {\bibinfo {volume} {79}},\ \bibinfo {pages} {011807} (\bibinfo {year} {2009})}\BibitemShut {NoStop}%
\bibitem [{\citenamefont {Bliokh}(2009)}]{bliokh2009geometrodynamics}%
  \BibitemOpen
  \bibfield  {author} {\bibinfo {author} {\bibfnamefont {K.~Y.}\ \bibnamefont {Bliokh}},\ }\href@noop {} {\bibfield  {journal} {\bibinfo  {journal} {Journal of Optics A: Pure and Applied Optics}\ }\textbf {\bibinfo {volume} {11}},\ \bibinfo {pages} {094009} (\bibinfo {year} {2009})}\BibitemShut {NoStop}%
\bibitem [{\citenamefont {Bliokh}(2006)}]{Bliokh_2006}%
  \BibitemOpen
  \bibfield  {author} {\bibinfo {author} {\bibfnamefont {K.~Y.}\ \bibnamefont {Bliokh}},\ }\href {https://doi.org/10.1103/PhysRevLett.97.043901} {\bibfield  {journal} {\bibinfo  {journal} {Phys. Rev. Lett.}\ }\textbf {\bibinfo {volume} {97}},\ \bibinfo {pages} {043901} (\bibinfo {year} {2006})}\BibitemShut {NoStop}%
\bibitem [{\citenamefont {Mieling}\ and\ \citenamefont {Oancea}(2023)}]{Mieling_2023}%
  \BibitemOpen
  \bibfield  {author} {\bibinfo {author} {\bibfnamefont {T.~B.}\ \bibnamefont {Mieling}}\ and\ \bibinfo {author} {\bibfnamefont {M.~A.}\ \bibnamefont {Oancea}},\ }\href {https://doi.org/10.1103/PhysRevResearch.5.023140} {\bibfield  {journal} {\bibinfo  {journal} {Phys. Rev. Res.}\ }\textbf {\bibinfo {volume} {5}},\ \bibinfo {pages} {023140} (\bibinfo {year} {2023})}\BibitemShut {NoStop}%
\bibitem [{\citenamefont {Fu}\ \emph {et~al.}(2019)\citenamefont {Fu}, \citenamefont {Guo}, \citenamefont {Liu}, \citenamefont {Li}, \citenamefont {Yin}, \citenamefont {Li},\ and\ \citenamefont {Chen}}]{fu2019spin}%
  \BibitemOpen
  \bibfield  {author} {\bibinfo {author} {\bibfnamefont {S.}~\bibnamefont {Fu}}, \bibinfo {author} {\bibfnamefont {C.}~\bibnamefont {Guo}}, \bibinfo {author} {\bibfnamefont {G.}~\bibnamefont {Liu}}, \bibinfo {author} {\bibfnamefont {Y.}~\bibnamefont {Li}}, \bibinfo {author} {\bibfnamefont {H.}~\bibnamefont {Yin}}, \bibinfo {author} {\bibfnamefont {Z.}~\bibnamefont {Li}},\ and\ \bibinfo {author} {\bibfnamefont {Z.}~\bibnamefont {Chen}},\ }\href@noop {} {\bibfield  {journal} {\bibinfo  {journal} {Physical Review Letters}\ }\textbf {\bibinfo {volume} {123}},\ \bibinfo {pages} {243904} (\bibinfo {year} {2019})}\BibitemShut {NoStop}%
\bibitem [{\citenamefont {Sheng}\ \emph {et~al.}(2023)\citenamefont {Sheng}, \citenamefont {Chen}, \citenamefont {Yuan}, \citenamefont {Liu}, \citenamefont {Zhang}, \citenamefont {Jing}, \citenamefont {Kuang},\ and\ \citenamefont {Zhou}}]{sheng2023photonic}%
  \BibitemOpen
  \bibfield  {author} {\bibinfo {author} {\bibfnamefont {L.}~\bibnamefont {Sheng}}, \bibinfo {author} {\bibfnamefont {Y.}~\bibnamefont {Chen}}, \bibinfo {author} {\bibfnamefont {S.}~\bibnamefont {Yuan}}, \bibinfo {author} {\bibfnamefont {X.}~\bibnamefont {Liu}}, \bibinfo {author} {\bibfnamefont {Z.}~\bibnamefont {Zhang}}, \bibinfo {author} {\bibfnamefont {H.}~\bibnamefont {Jing}}, \bibinfo {author} {\bibfnamefont {L.-M.}\ \bibnamefont {Kuang}},\ and\ \bibinfo {author} {\bibfnamefont {X.}~\bibnamefont {Zhou}},\ }\href@noop {} {\bibfield  {journal} {\bibinfo  {journal} {Progress in Quantum Electronics}\ ,\ \bibinfo {pages} {100484}} (\bibinfo {year} {2023})}\BibitemShut {NoStop}%
\bibitem [{\citenamefont {Ling}\ \emph {et~al.}(2017)\citenamefont {Ling}, \citenamefont {Zhou}, \citenamefont {Huang}, \citenamefont {Liu}, \citenamefont {Qiu}, \citenamefont {Luo},\ and\ \citenamefont {Wen}}]{ling2017recent}%
  \BibitemOpen
  \bibfield  {author} {\bibinfo {author} {\bibfnamefont {X.}~\bibnamefont {Ling}}, \bibinfo {author} {\bibfnamefont {X.}~\bibnamefont {Zhou}}, \bibinfo {author} {\bibfnamefont {K.}~\bibnamefont {Huang}}, \bibinfo {author} {\bibfnamefont {Y.}~\bibnamefont {Liu}}, \bibinfo {author} {\bibfnamefont {C.-W.}\ \bibnamefont {Qiu}}, \bibinfo {author} {\bibfnamefont {H.}~\bibnamefont {Luo}},\ and\ \bibinfo {author} {\bibfnamefont {S.}~\bibnamefont {Wen}},\ }\href@noop {} {\bibfield  {journal} {\bibinfo  {journal} {Reports on Progress in Physics}\ }\textbf {\bibinfo {volume} {80}},\ \bibinfo {pages} {066401} (\bibinfo {year} {2017})}\BibitemShut {NoStop}%
\bibitem [{\citenamefont {Hosten}\ and\ \citenamefont {Kwiat}(2008)}]{hosten2008observation}%
  \BibitemOpen
  \bibfield  {author} {\bibinfo {author} {\bibfnamefont {O.}~\bibnamefont {Hosten}}\ and\ \bibinfo {author} {\bibfnamefont {P.}~\bibnamefont {Kwiat}},\ }\href@noop {} {\bibfield  {journal} {\bibinfo  {journal} {Science}\ }\textbf {\bibinfo {volume} {319}},\ \bibinfo {pages} {787} (\bibinfo {year} {2008})}\BibitemShut {NoStop}%
\bibitem [{\citenamefont {Bliokh}\ \emph {et~al.}(2008)\citenamefont {Bliokh}, \citenamefont {Niv}, \citenamefont {Kleiner},\ and\ \citenamefont {Hasman}}]{bliokh2008geometrodynamics}%
  \BibitemOpen
  \bibfield  {author} {\bibinfo {author} {\bibfnamefont {K.~Y.}\ \bibnamefont {Bliokh}}, \bibinfo {author} {\bibfnamefont {A.}~\bibnamefont {Niv}}, \bibinfo {author} {\bibfnamefont {V.}~\bibnamefont {Kleiner}},\ and\ \bibinfo {author} {\bibfnamefont {E.}~\bibnamefont {Hasman}},\ }\href@noop {} {\bibfield  {journal} {\bibinfo  {journal} {Nature Photonics}\ }\textbf {\bibinfo {volume} {2}},\ \bibinfo {pages} {748} (\bibinfo {year} {2008})}\BibitemShut {NoStop}%
\bibitem [{\citenamefont {Bliokh}\ \emph {et~al.}(2015)\citenamefont {Bliokh}, \citenamefont {Rodr\'\i{}guez-Fortu\~no}, \citenamefont {Nori},\ and\ \citenamefont {Zayats}}]{Bliokh_2015}%
  \BibitemOpen
  \bibfield  {author} {\bibinfo {author} {\bibfnamefont {K.~Y.}\ \bibnamefont {Bliokh}}, \bibinfo {author} {\bibfnamefont {F.~J.}\ \bibnamefont {Rodr\'\i{}guez-Fortu\~no}}, \bibinfo {author} {\bibfnamefont {F.}~\bibnamefont {Nori}},\ and\ \bibinfo {author} {\bibfnamefont {A.~V.}\ \bibnamefont {Zayats}},\ }\href {https://doi.org/10.1038/nphoton.2015.201} {\bibfield  {journal} {\bibinfo  {journal} {Nature Photon.}\ }\textbf {\bibinfo {volume} {9}},\ \bibinfo {pages} {796} (\bibinfo {year} {2015})}\BibitemShut {NoStop}%
\bibitem [{\citenamefont {Luo}\ \emph {et~al.}(2011)\citenamefont {Luo}, \citenamefont {Zhou}, \citenamefont {Shu}, \citenamefont {Wen},\ and\ \citenamefont {Fan}}]{PhysRevA.84.043806}%
  \BibitemOpen
  \bibfield  {author} {\bibinfo {author} {\bibfnamefont {H.}~\bibnamefont {Luo}}, \bibinfo {author} {\bibfnamefont {X.}~\bibnamefont {Zhou}}, \bibinfo {author} {\bibfnamefont {W.}~\bibnamefont {Shu}}, \bibinfo {author} {\bibfnamefont {S.}~\bibnamefont {Wen}},\ and\ \bibinfo {author} {\bibfnamefont {D.}~\bibnamefont {Fan}},\ }\href {https://doi.org/10.1103/PhysRevA.84.043806} {\bibfield  {journal} {\bibinfo  {journal} {Phys. Rev. A}\ }\textbf {\bibinfo {volume} {84}},\ \bibinfo {pages} {043806} (\bibinfo {year} {2011})}\BibitemShut {NoStop}%
\bibitem [{\citenamefont {Haefner}\ \emph {et~al.}(2009)\citenamefont {Haefner}, \citenamefont {Sukhov},\ and\ \citenamefont {Dogariu}}]{PhysRevLett.102.123903}%
  \BibitemOpen
  \bibfield  {author} {\bibinfo {author} {\bibfnamefont {D.}~\bibnamefont {Haefner}}, \bibinfo {author} {\bibfnamefont {S.}~\bibnamefont {Sukhov}},\ and\ \bibinfo {author} {\bibfnamefont {A.}~\bibnamefont {Dogariu}},\ }\href@noop {} {\bibfield  {journal} {\bibinfo  {journal} {Phys. Rev. Lett.}\ }\textbf {\bibinfo {volume} {102}},\ \bibinfo {pages} {123903} (\bibinfo {year} {2009})}\BibitemShut {NoStop}%
\bibitem [{\citenamefont {Duval}\ \emph {et~al.}(2019)\citenamefont {Duval}, \citenamefont {Marsot},\ and\ \citenamefont {Sch\"ucker}}]{Duval:2018hzh}%
  \BibitemOpen
  \bibfield  {author} {\bibinfo {author} {\bibfnamefont {C.}~\bibnamefont {Duval}}, \bibinfo {author} {\bibfnamefont {L.}~\bibnamefont {Marsot}},\ and\ \bibinfo {author} {\bibfnamefont {T.}~\bibnamefont {Sch\"ucker}},\ }\href {https://doi.org/10.1103/PhysRevD.99.124037} {\bibfield  {journal} {\bibinfo  {journal} {Phys. Rev. D}\ }\textbf {\bibinfo {volume} {99}},\ \bibinfo {pages} {124037} (\bibinfo {year} {2019})},\ \Eprint {https://arxiv.org/abs/1812.03014} {arXiv:1812.03014 [gr-qc]} \BibitemShut {NoStop}%
\bibitem [{\citenamefont {Duval}\ \emph {et~al.}(2013)\citenamefont {Duval}, \citenamefont {Horvathy},\ and\ \citenamefont {Zhang}}]{Duval:2012ye}%
  \BibitemOpen
  \bibfield  {author} {\bibinfo {author} {\bibfnamefont {C.}~\bibnamefont {Duval}}, \bibinfo {author} {\bibfnamefont {P.~A.}\ \bibnamefont {Horvathy}},\ and\ \bibinfo {author} {\bibfnamefont {P.~M.}\ \bibnamefont {Zhang}},\ }\href {https://doi.org/10.1088/2040-8978/15/1/014005} {\bibfield  {journal} {\bibinfo  {journal} {J. Opt.}\ }\textbf {\bibinfo {volume} {15}},\ \bibinfo {pages} {014005} (\bibinfo {year} {2013})},\ \Eprint {https://arxiv.org/abs/1202.2430} {arXiv:1202.2430 [physics.optics]} \BibitemShut {NoStop}%
\bibitem [{\citenamefont {Duval}\ \emph {et~al.}(2007)\citenamefont {Duval}, \citenamefont {Horvath},\ and\ \citenamefont {Horvathy}}]{Duval:2005ry}%
  \BibitemOpen
  \bibfield  {author} {\bibinfo {author} {\bibfnamefont {C.}~\bibnamefont {Duval}}, \bibinfo {author} {\bibfnamefont {Z.}~\bibnamefont {Horvath}},\ and\ \bibinfo {author} {\bibfnamefont {P.}~\bibnamefont {Horvathy}},\ }\href {https://doi.org/10.1016/j.geomphys.2006.07.003} {\bibfield  {journal} {\bibinfo  {journal} {J. Geom. Phys.}\ }\textbf {\bibinfo {volume} {57}},\ \bibinfo {pages} {925} (\bibinfo {year} {2007})},\ \Eprint {https://arxiv.org/abs/math-ph/0509031} {arXiv:math-ph/0509031} \BibitemShut {NoStop}%
\bibitem [{\citenamefont {Duval}\ \emph {et~al.}(2006)\citenamefont {Duval}, \citenamefont {Horvath},\ and\ \citenamefont {Horvathy}}]{Duval:2005ky}%
  \BibitemOpen
  \bibfield  {author} {\bibinfo {author} {\bibfnamefont {C.}~\bibnamefont {Duval}}, \bibinfo {author} {\bibfnamefont {Z.}~\bibnamefont {Horvath}},\ and\ \bibinfo {author} {\bibfnamefont {P.~A.}\ \bibnamefont {Horvathy}},\ }\href {https://doi.org/10.1103/PhysRevD.74.021701} {\bibfield  {journal} {\bibinfo  {journal} {Phys. Rev. D}\ }\textbf {\bibinfo {volume} {74}},\ \bibinfo {pages} {021701} (\bibinfo {year} {2006})},\ \Eprint {https://arxiv.org/abs/cond-mat/0509636} {arXiv:cond-mat/0509636} \BibitemShut {NoStop}%
\bibitem [{\citenamefont {Beijersbergen}\ \emph {et~al.}(1993)\citenamefont {Beijersbergen}, \citenamefont {Allen}, \citenamefont {Van~der Veen},\ and\ \citenamefont {Woerdman}}]{beijersbergen1993astigmatic}%
  \BibitemOpen
  \bibfield  {author} {\bibinfo {author} {\bibfnamefont {M.~W.}\ \bibnamefont {Beijersbergen}}, \bibinfo {author} {\bibfnamefont {L.}~\bibnamefont {Allen}}, \bibinfo {author} {\bibfnamefont {H.}~\bibnamefont {Van~der Veen}},\ and\ \bibinfo {author} {\bibfnamefont {J.}~\bibnamefont {Woerdman}},\ }\href@noop {} {\bibfield  {journal} {\bibinfo  {journal} {Optics Communications}\ }\textbf {\bibinfo {volume} {96}},\ \bibinfo {pages} {123} (\bibinfo {year} {1993})}\BibitemShut {NoStop}%
\bibitem [{\citenamefont {Bliokh}\ \emph {et~al.}(2017)\citenamefont {Bliokh}, \citenamefont {Ivanov}, \citenamefont {Guzzinati}, \citenamefont {Clark}, \citenamefont {Van~Boxem}, \citenamefont {B{\'e}ch{\'e}}, \citenamefont {Juchtmans}, \citenamefont {Alonso}, \citenamefont {Schattschneider}, \citenamefont {Nori} \emph {et~al.}}]{bliokh2017theory}%
  \BibitemOpen
  \bibfield  {author} {\bibinfo {author} {\bibfnamefont {K.~Y.}\ \bibnamefont {Bliokh}}, \bibinfo {author} {\bibfnamefont {I.~P.}\ \bibnamefont {Ivanov}}, \bibinfo {author} {\bibfnamefont {G.}~\bibnamefont {Guzzinati}}, \bibinfo {author} {\bibfnamefont {L.}~\bibnamefont {Clark}}, \bibinfo {author} {\bibfnamefont {R.}~\bibnamefont {Van~Boxem}}, \bibinfo {author} {\bibfnamefont {A.}~\bibnamefont {B{\'e}ch{\'e}}}, \bibinfo {author} {\bibfnamefont {R.}~\bibnamefont {Juchtmans}}, \bibinfo {author} {\bibfnamefont {M.~A.}\ \bibnamefont {Alonso}}, \bibinfo {author} {\bibfnamefont {P.}~\bibnamefont {Schattschneider}}, \bibinfo {author} {\bibfnamefont {F.}~\bibnamefont {Nori}}, \emph {et~al.},\ }\href@noop {} {\bibfield  {journal} {\bibinfo  {journal} {Physics Reports}\ }\textbf {\bibinfo {volume} {690}},\ \bibinfo {pages} {1} (\bibinfo {year} {2017})}\BibitemShut {NoStop}%
\bibitem [{\citenamefont {Lai}\ \emph {et~al.}(2018)\citenamefont {Lai}, \citenamefont {Wang}, \citenamefont {Liang}, \citenamefont {Wang},\ and\ \citenamefont {Zong}}]{PhysRevA.97.033843}%
  \BibitemOpen
  \bibfield  {author} {\bibinfo {author} {\bibfnamefont {M.-Y.}\ \bibnamefont {Lai}}, \bibinfo {author} {\bibfnamefont {Y.-L.}\ \bibnamefont {Wang}}, \bibinfo {author} {\bibfnamefont {G.-H.}\ \bibnamefont {Liang}}, \bibinfo {author} {\bibfnamefont {F.}~\bibnamefont {Wang}},\ and\ \bibinfo {author} {\bibfnamefont {H.-S.}\ \bibnamefont {Zong}},\ }\href {https://doi.org/10.1103/PhysRevA.97.033843} {\bibfield  {journal} {\bibinfo  {journal} {Phys. Rev. A}\ }\textbf {\bibinfo {volume} {97}},\ \bibinfo {pages} {033843} (\bibinfo {year} {2018})}\BibitemShut {NoStop}%
\bibitem [{\citenamefont {Ruiz}\ and\ \citenamefont {Dodin}(2015)}]{PhysRevA.92.043805}%
  \BibitemOpen
  \bibfield  {author} {\bibinfo {author} {\bibfnamefont {D.~E.}\ \bibnamefont {Ruiz}}\ and\ \bibinfo {author} {\bibfnamefont {I.~Y.}\ \bibnamefont {Dodin}},\ }\href@noop {} {\bibfield  {journal} {\bibinfo  {journal} {Phys. Rev. A}\ }\textbf {\bibinfo {volume} {92}},\ \bibinfo {pages} {043805} (\bibinfo {year} {2015})}\BibitemShut {NoStop}%
\bibitem [{\citenamefont {Wang}\ and\ \citenamefont {Chen}(2019)}]{wang2019anomalous}%
  \BibitemOpen
  \bibfield  {author} {\bibinfo {author} {\bibfnamefont {Z.-L.}\ \bibnamefont {Wang}}\ and\ \bibinfo {author} {\bibfnamefont {X.-S.}\ \bibnamefont {Chen}},\ }\href@noop {} {\bibfield  {journal} {\bibinfo  {journal} {Physical Review A}\ }\textbf {\bibinfo {volume} {99}},\ \bibinfo {pages} {063832} (\bibinfo {year} {2019})}\BibitemShut {NoStop}%
\bibitem [{\citenamefont {Choi}\ \emph {et~al.}(2023)\citenamefont {Choi}, \citenamefont {Jo}, \citenamefont {Ko}, \citenamefont {Go}, \citenamefont {Kim}, \citenamefont {Park}, \citenamefont {Kim}, \citenamefont {Min}, \citenamefont {Choi},\ and\ \citenamefont {Lee}}]{choi2023observation}%
  \BibitemOpen
  \bibfield  {author} {\bibinfo {author} {\bibfnamefont {Y.-G.}\ \bibnamefont {Choi}}, \bibinfo {author} {\bibfnamefont {D.}~\bibnamefont {Jo}}, \bibinfo {author} {\bibfnamefont {K.-H.}\ \bibnamefont {Ko}}, \bibinfo {author} {\bibfnamefont {D.}~\bibnamefont {Go}}, \bibinfo {author} {\bibfnamefont {K.-H.}\ \bibnamefont {Kim}}, \bibinfo {author} {\bibfnamefont {H.~G.}\ \bibnamefont {Park}}, \bibinfo {author} {\bibfnamefont {C.}~\bibnamefont {Kim}}, \bibinfo {author} {\bibfnamefont {B.-C.}\ \bibnamefont {Min}}, \bibinfo {author} {\bibfnamefont {G.-M.}\ \bibnamefont {Choi}},\ and\ \bibinfo {author} {\bibfnamefont {H.-W.}\ \bibnamefont {Lee}},\ }\href@noop {} {\bibfield  {journal} {\bibinfo  {journal} {Nature}\ }\textbf {\bibinfo {volume} {619}},\ \bibinfo {pages} {52} (\bibinfo {year} {2023})}\BibitemShut {NoStop}%
\bibitem [{\citenamefont {Qiu}\ \emph {et~al.}(2024)\citenamefont {Qiu}, \citenamefont {Lian},\ and\ \citenamefont {Zhang}}]{Qiu:2024rsr}%
  \BibitemOpen
  \bibfield  {author} {\bibinfo {author} {\bibfnamefont {W.-S.}\ \bibnamefont {Qiu}}, \bibinfo {author} {\bibfnamefont {D.-D.}\ \bibnamefont {Lian}},\ and\ \bibinfo {author} {\bibfnamefont {P.-M.}\ \bibnamefont {Zhang}},\ }\href {https://doi.org/10.1140/epjc/s10052-024-13409-x} {\bibfield  {journal} {\bibinfo  {journal} {Eur. Phys. J. C}\ }\textbf {\bibinfo {volume} {84}},\ \bibinfo {pages} {1013} (\bibinfo {year} {2024})},\ \Eprint {https://arxiv.org/abs/2407.06553} {arXiv:2407.06553 [gr-qc]} \BibitemShut {NoStop}%
\bibitem [{\citenamefont {Bialynicki-Birula}(1996)}]{BIALYNICKIBIRULA1996245}%
  \BibitemOpen
  \bibfield  {author} {\bibinfo {author} {\bibfnamefont {I.}~\bibnamefont {Bialynicki-Birula}}\ }(\bibinfo  {publisher} {Elsevier},\ \bibinfo {year} {1996})\ pp.\ \bibinfo {pages} {245--294}\BibitemShut {NoStop}%
\bibitem [{\citenamefont {{Narimanov}}\ and\ \citenamefont {{Kildishev}}(2009)}]{narimanov2009optical}%
  \BibitemOpen
  \bibfield  {author} {\bibinfo {author} {\bibfnamefont {E.~E.}\ \bibnamefont {{Narimanov}}}\ and\ \bibinfo {author} {\bibfnamefont {A.~V.}\ \bibnamefont {{Kildishev}}},\ }\href {https://doi.org/10.1063/1.3184594} {\bibfield  {journal} {\bibinfo  {journal} {Applied Physics Letters}\ }\textbf {\bibinfo {volume} {95}},\ \bibinfo {pages} {041106} (\bibinfo {year} {2009})}\BibitemShut {NoStop}%
\bibitem [{\citenamefont {Genov}\ \emph {et~al.}(2009)\citenamefont {Genov}, \citenamefont {Zhang},\ and\ \citenamefont {Zhang}}]{genov2009mimicking}%
  \BibitemOpen
  \bibfield  {author} {\bibinfo {author} {\bibfnamefont {D.~A.}\ \bibnamefont {Genov}}, \bibinfo {author} {\bibfnamefont {S.}~\bibnamefont {Zhang}},\ and\ \bibinfo {author} {\bibfnamefont {X.}~\bibnamefont {Zhang}},\ }\href@noop {} {\bibfield  {journal} {\bibinfo  {journal} {Nature Physics}\ }\textbf {\bibinfo {volume} {5}},\ \bibinfo {pages} {687} (\bibinfo {year} {2009})}\BibitemShut {NoStop}%
\bibitem [{\citenamefont {Cheng}\ \emph {et~al.}(2010)\citenamefont {Cheng}, \citenamefont {Cui}, \citenamefont {Jiang},\ and\ \citenamefont {Cai}}]{cheng2010omnidirectional}%
  \BibitemOpen
  \bibfield  {author} {\bibinfo {author} {\bibfnamefont {Q.}~\bibnamefont {Cheng}}, \bibinfo {author} {\bibfnamefont {T.~J.}\ \bibnamefont {Cui}}, \bibinfo {author} {\bibfnamefont {W.~X.}\ \bibnamefont {Jiang}},\ and\ \bibinfo {author} {\bibfnamefont {B.~G.}\ \bibnamefont {Cai}},\ }\href@noop {} {\bibfield  {journal} {\bibinfo  {journal} {New Journal of Physics}\ }\textbf {\bibinfo {volume} {12}},\ \bibinfo {pages} {063006} (\bibinfo {year} {2010})}\BibitemShut {NoStop}%
\bibitem [{\citenamefont {Yang}\ \emph {et~al.}(2012)\citenamefont {Yang}, \citenamefont {Leng}, \citenamefont {Wang}, \citenamefont {Ma},\ and\ \citenamefont {Ong}}]{yang2012electromagnetic}%
  \BibitemOpen
  \bibfield  {author} {\bibinfo {author} {\bibfnamefont {Y.}~\bibnamefont {Yang}}, \bibinfo {author} {\bibfnamefont {L.~Y.}\ \bibnamefont {Leng}}, \bibinfo {author} {\bibfnamefont {N.}~\bibnamefont {Wang}}, \bibinfo {author} {\bibfnamefont {Y.}~\bibnamefont {Ma}},\ and\ \bibinfo {author} {\bibfnamefont {C.}~\bibnamefont {Ong}},\ }\href@noop {} {\bibfield  {journal} {\bibinfo  {journal} {JOSA A}\ }\textbf {\bibinfo {volume} {29}},\ \bibinfo {pages} {473} (\bibinfo {year} {2012})}\BibitemShut {NoStop}%
\bibitem [{\citenamefont {Sheng}\ \emph {et~al.}(2013)\citenamefont {Sheng}, \citenamefont {Liu}, \citenamefont {Wang}, \citenamefont {Zhu},\ and\ \citenamefont {Genov}}]{sheng2013trapping}%
  \BibitemOpen
  \bibfield  {author} {\bibinfo {author} {\bibfnamefont {C.}~\bibnamefont {Sheng}}, \bibinfo {author} {\bibfnamefont {H.}~\bibnamefont {Liu}}, \bibinfo {author} {\bibfnamefont {Y.}~\bibnamefont {Wang}}, \bibinfo {author} {\bibfnamefont {S.}~\bibnamefont {Zhu}},\ and\ \bibinfo {author} {\bibfnamefont {D.}~\bibnamefont {Genov}},\ }\href@noop {} {\bibfield  {journal} {\bibinfo  {journal} {Nature Photonics}\ }\textbf {\bibinfo {volume} {7}},\ \bibinfo {pages} {902} (\bibinfo {year} {2013})}\BibitemShut {NoStop}%
\bibitem [{\citenamefont {Lian}\ and\ \citenamefont {Chen}(2022)}]{PhysRevD.105.104008}%
  \BibitemOpen
  \bibfield  {author} {\bibinfo {author} {\bibfnamefont {D.-D.}\ \bibnamefont {Lian}}\ and\ \bibinfo {author} {\bibfnamefont {X.-S.}\ \bibnamefont {Chen}},\ }\href {https://doi.org/10.1103/PhysRevD.105.104008} {\bibfield  {journal} {\bibinfo  {journal} {Phys. Rev. D}\ }\textbf {\bibinfo {volume} {105}},\ \bibinfo {pages} {104008} (\bibinfo {year} {2022})}\BibitemShut {NoStop}%
\bibitem [{\citenamefont {Lian}\ and\ \citenamefont {Zhang}(2024)}]{lian2024motion}%
  \BibitemOpen
  \bibfield  {author} {\bibinfo {author} {\bibfnamefont {D.-D.}\ \bibnamefont {Lian}}\ and\ \bibinfo {author} {\bibfnamefont {P.-M.}\ \bibnamefont {Zhang}},\ }\href {https://doi.org/10.1088/1361-6382/ad721d} {\bibfield  {journal} {\bibinfo  {journal} {Class. Quant. Grav.}\ }\textbf {\bibinfo {volume} {41}},\ \bibinfo {pages} {195007} (\bibinfo {year} {2024})},\ \Eprint {https://arxiv.org/abs/2312.14391} {arXiv:2312.14391 [gr-qc]} \BibitemShut {NoStop}%
\bibitem [{\citenamefont {Plebanski}(1960)}]{PhysRev.118.1396}%
  \BibitemOpen
  \bibfield  {author} {\bibinfo {author} {\bibfnamefont {J.}~\bibnamefont {Plebanski}},\ }\href {https://doi.org/10.1103/PhysRev.118.1396} {\bibfield  {journal} {\bibinfo  {journal} {Phys. Rev.}\ }\textbf {\bibinfo {volume} {118}},\ \bibinfo {pages} {1396} (\bibinfo {year} {1960})}\BibitemShut {NoStop}%
\bibitem [{\citenamefont {Carini}\ \emph {et~al.}(1992)\citenamefont {Carini}, \citenamefont {Feng}, \citenamefont {Li},\ and\ \citenamefont {Ruffini}}]{carini1992phase}%
  \BibitemOpen
  \bibfield  {author} {\bibinfo {author} {\bibfnamefont {P.}~\bibnamefont {Carini}}, \bibinfo {author} {\bibfnamefont {L.~L.}\ \bibnamefont {Feng}}, \bibinfo {author} {\bibfnamefont {M.}~\bibnamefont {Li}},\ and\ \bibinfo {author} {\bibfnamefont {R.}~\bibnamefont {Ruffini}},\ }\href@noop {} {\bibfield  {journal} {\bibinfo  {journal} {Physical Review D}\ }\textbf {\bibinfo {volume} {46}},\ \bibinfo {pages} {5407} (\bibinfo {year} {1992})}\BibitemShut {NoStop}%
\bibitem [{\citenamefont {Wu}\ \emph {et~al.}(2022)\citenamefont {Wu}, \citenamefont {Zhu},\ and\ \citenamefont {Feng}}]{wu2022testing}%
  \BibitemOpen
  \bibfield  {author} {\bibinfo {author} {\bibfnamefont {Q.}~\bibnamefont {Wu}}, \bibinfo {author} {\bibfnamefont {W.}~\bibnamefont {Zhu}},\ and\ \bibinfo {author} {\bibfnamefont {L.}~\bibnamefont {Feng}},\ }\href@noop {} {\bibfield  {journal} {\bibinfo  {journal} {Universe}\ }\textbf {\bibinfo {volume} {8}},\ \bibinfo {pages} {535} (\bibinfo {year} {2022})}\BibitemShut {NoStop}%
\bibitem [{\citenamefont {Chanda}\ \emph {et~al.}(2019)\citenamefont {Chanda}, \citenamefont {Gibbons}, \citenamefont {Guha}, \citenamefont {Maraner},\ and\ \citenamefont {Werner}}]{chanda2019jacobi}%
  \BibitemOpen
  \bibfield  {author} {\bibinfo {author} {\bibfnamefont {S.}~\bibnamefont {Chanda}}, \bibinfo {author} {\bibfnamefont {G.}~\bibnamefont {Gibbons}}, \bibinfo {author} {\bibfnamefont {P.}~\bibnamefont {Guha}}, \bibinfo {author} {\bibfnamefont {P.}~\bibnamefont {Maraner}},\ and\ \bibinfo {author} {\bibfnamefont {M.~C.}\ \bibnamefont {Werner}},\ }\href@noop {} {\bibfield  {journal} {\bibinfo  {journal} {Journal of Mathematical Physics}\ }\textbf {\bibinfo {volume} {60}} (\bibinfo {year} {2019})}\BibitemShut {NoStop}%
\bibitem [{\citenamefont {Gibbons}\ and\ \citenamefont {Werner}(2019)}]{gibbons2019gravitational}%
  \BibitemOpen
  \bibfield  {author} {\bibinfo {author} {\bibfnamefont {G.~W.}\ \bibnamefont {Gibbons}}\ and\ \bibinfo {author} {\bibfnamefont {M.~C.}\ \bibnamefont {Werner}},\ }\href@noop {} {\bibfield  {journal} {\bibinfo  {journal} {Universe}\ }\textbf {\bibinfo {volume} {5}},\ \bibinfo {pages} {88} (\bibinfo {year} {2019})}\BibitemShut {NoStop}%
\bibitem [{\citenamefont {Fern{\'a}ndez-N{\'u}{\~n}ez}\ and\ \citenamefont {Bulashenko}(2016)}]{fernandez2016anisotropic}%
  \BibitemOpen
  \bibfield  {author} {\bibinfo {author} {\bibfnamefont {I.}~\bibnamefont {Fern{\'a}ndez-N{\'u}{\~n}ez}}\ and\ \bibinfo {author} {\bibfnamefont {O.}~\bibnamefont {Bulashenko}},\ }\href@noop {} {\bibfield  {journal} {\bibinfo  {journal} {Physics Letters A}\ }\textbf {\bibinfo {volume} {380}},\ \bibinfo {pages} {1} (\bibinfo {year} {2016})}\BibitemShut {NoStop}%
\bibitem [{\citenamefont {Thompson}(2018)}]{PhysRevD.97.065001}%
  \BibitemOpen
  \bibfield  {author} {\bibinfo {author} {\bibfnamefont {R.~T.}\ \bibnamefont {Thompson}},\ }\href {https://doi.org/10.1103/PhysRevD.97.065001} {\bibfield  {journal} {\bibinfo  {journal} {Phys. Rev. D}\ }\textbf {\bibinfo {volume} {97}},\ \bibinfo {pages} {065001} (\bibinfo {year} {2018})}\BibitemShut {NoStop}%
\bibitem [{\citenamefont {Gordon}(1923)}]{gordon1923lichtfortpflanzung}%
  \BibitemOpen
  \bibfield  {author} {\bibinfo {author} {\bibfnamefont {W.}~\bibnamefont {Gordon}},\ }\href@noop {} {\bibfield  {journal} {\bibinfo  {journal} {Annalen der Physik}\ }\textbf {\bibinfo {volume} {377}},\ \bibinfo {pages} {421} (\bibinfo {year} {1923})}\BibitemShut {NoStop}%
\bibitem [{\citenamefont {Harte}\ and\ \citenamefont {Oancea}(2022)}]{PhysRevD.105.104061}%
  \BibitemOpen
  \bibfield  {author} {\bibinfo {author} {\bibfnamefont {A.~I.}\ \bibnamefont {Harte}}\ and\ \bibinfo {author} {\bibfnamefont {M.~A.}\ \bibnamefont {Oancea}},\ }\href {https://doi.org/10.1103/PhysRevD.105.104061} {\bibfield  {journal} {\bibinfo  {journal} {Phys. Rev. D}\ }\textbf {\bibinfo {volume} {105}},\ \bibinfo {pages} {104061} (\bibinfo {year} {2022})}\BibitemShut {NoStop}%
\bibitem [{\citenamefont {Liu}\ and\ \citenamefont {Ivanov}(2023)}]{liu2023threshold}%
  \BibitemOpen
  \bibfield  {author} {\bibinfo {author} {\bibfnamefont {B.}~\bibnamefont {Liu}}\ and\ \bibinfo {author} {\bibfnamefont {I.~P.}\ \bibnamefont {Ivanov}},\ }\href@noop {} {\bibfield  {journal} {\bibinfo  {journal} {Physical Review A}\ }\textbf {\bibinfo {volume} {107}},\ \bibinfo {pages} {063110} (\bibinfo {year} {2023})}\BibitemShut {NoStop}%
\bibitem [{\citenamefont {Peskin}(2018)}]{peskin2018introduction}%
  \BibitemOpen
  \bibfield  {author} {\bibinfo {author} {\bibfnamefont {M.~E.}\ \bibnamefont {Peskin}},\ }\href@noop {} {\emph {\bibinfo {title} {An introduction to quantum field theory}}}\ (\bibinfo  {publisher} {CRC press},\ \bibinfo {year} {2018})\BibitemShut {NoStop}%
\bibitem [{\citenamefont {Sheng}\ \emph {et~al.}(2016)\citenamefont {Sheng}, \citenamefont {Bekenstein}, \citenamefont {Liu}, \citenamefont {Zhu},\ and\ \citenamefont {Segev}}]{sheng2016wavefront}%
  \BibitemOpen
  \bibfield  {author} {\bibinfo {author} {\bibfnamefont {C.}~\bibnamefont {Sheng}}, \bibinfo {author} {\bibfnamefont {R.}~\bibnamefont {Bekenstein}}, \bibinfo {author} {\bibfnamefont {H.}~\bibnamefont {Liu}}, \bibinfo {author} {\bibfnamefont {S.}~\bibnamefont {Zhu}},\ and\ \bibinfo {author} {\bibfnamefont {M.}~\bibnamefont {Segev}},\ }\href@noop {} {\bibfield  {journal} {\bibinfo  {journal} {Nature communications}\ }\textbf {\bibinfo {volume} {7}},\ \bibinfo {pages} {10747} (\bibinfo {year} {2016})}\BibitemShut {NoStop}%
\bibitem [{\citenamefont {Wang}\ \emph {et~al.}(2021)\citenamefont {Wang}, \citenamefont {Tan}, \citenamefont {Liang}, \citenamefont {Ma}, \citenamefont {Wang},\ and\ \citenamefont {Cheng}}]{wang2021generalized}%
  \BibitemOpen
  \bibfield  {author} {\bibinfo {author} {\bibfnamefont {W.}~\bibnamefont {Wang}}, \bibinfo {author} {\bibfnamefont {Y.}~\bibnamefont {Tan}}, \bibinfo {author} {\bibfnamefont {B.}~\bibnamefont {Liang}}, \bibinfo {author} {\bibfnamefont {G.}~\bibnamefont {Ma}}, \bibinfo {author} {\bibfnamefont {S.}~\bibnamefont {Wang}},\ and\ \bibinfo {author} {\bibfnamefont {J.}~\bibnamefont {Cheng}},\ }\href@noop {} {\bibfield  {journal} {\bibinfo  {journal} {Physical Review B}\ }\textbf {\bibinfo {volume} {104}},\ \bibinfo {pages} {174301} (\bibinfo {year} {2021})}\BibitemShut {NoStop}%
\bibitem [{\citenamefont {Dasgupta}\ and\ \citenamefont {Gupta}(2005)}]{dasgupta2005experimental}%
  \BibitemOpen
  \bibfield  {author} {\bibinfo {author} {\bibfnamefont {R.}~\bibnamefont {Dasgupta}}\ and\ \bibinfo {author} {\bibfnamefont {P.}~\bibnamefont {Gupta}},\ }in\ \href@noop {} {\emph {\bibinfo {booktitle} {Conference on Lasers and Electro-Optics}}}\ (\bibinfo {organization} {Optica Publishing Group},\ \bibinfo {year} {2005})\ p.\ \bibinfo {pages} {CThBB2}\BibitemShut {NoStop}%
\bibitem [{\citenamefont {Courtial}\ \emph {et~al.}(1997)\citenamefont {Courtial}, \citenamefont {Dholakia}, \citenamefont {Allen},\ and\ \citenamefont {Padgett}}]{courtial1997gaussian}%
  \BibitemOpen
  \bibfield  {author} {\bibinfo {author} {\bibfnamefont {J.}~\bibnamefont {Courtial}}, \bibinfo {author} {\bibfnamefont {K.}~\bibnamefont {Dholakia}}, \bibinfo {author} {\bibfnamefont {L.}~\bibnamefont {Allen}},\ and\ \bibinfo {author} {\bibfnamefont {M.}~\bibnamefont {Padgett}},\ }\href@noop {} {\bibfield  {journal} {\bibinfo  {journal} {Optics communications}\ }\textbf {\bibinfo {volume} {144}},\ \bibinfo {pages} {210} (\bibinfo {year} {1997})}\BibitemShut {NoStop}%
\end{thebibliography}%
\end{document}